\documentclass[trackchanges]{aastex701}

\graphicspath{{./}{Figures/}}
\usepackage{amsmath}
\usepackage{subfigure}

\begin{document}

\title{Age of the Pisces-Eridanus Stellar Stream: \\ Binary Star $\tau^1$ Aqr and Other Bright Members}

\author[0000-0001-7237-3787]{Jynessa Valladon}
\affiliation{San Diego State University, Department of Astronomy, San Diego, CA 92182, USA}
\email{jvalladon@sdsu.edu}
 
\author[0000-0003-4070-4881]{Eric L. Sandquist} 
\affiliation{San Diego State University, Department of Astronomy, San Diego, CA 92182, USA}
\email{esandquist@sdsu.edu}

\author[0000-0001-5415-9189]{Gail H. Schaefer} 
\affiliation{The CHARA Array of Georgia State University, Mount Wilson Observatory, Mount Wilson, CA 91023, USA}
\email{gschaefer@gsu.edu}

\author[0000-0001-9745-5834]{Cyprien Lanthermann} 
\affiliation{The CHARA Array of Georgia State University, Mount Wilson Observatory, Mount Wilson, CA 91023, USA}
\email{clanthermann@gsu.edu}

\author[0000-0003-3045-5148]{Jeremy Jones} 
\affiliation{The CHARA Array of Georgia State University, Mount Wilson Observatory, Mount Wilson, CA 91023, USA}
\email{jjones176@gsu.edu}

\author[0009-0005-8004-2351]{Becky Flores} 
\affiliation{The CHARA Array of Georgia State University, Mount Wilson Observatory, Mount Wilson, CA 91023, USA}
\email{bflores5@gsu.edu}

\author[0000-0001-9647-2886]{Jerome A. Orosz} 
\affiliation{San Diego State University, Department of Astronomy, San Diego, CA 92182, USA}
\email{jorosz@sdsu.edu}

\author[0000-0002-5379-1286]{Markus Scholler} 
\affiliation{European Southern Observatory, Karl-Schwarzschild-Strasse, 2, 85748 Garching-bei-München,
Germany}
\email{mschoell@eso.org}

\author[0000-0002-4308-0763]{Christian Hummel} 
\affiliation{European Southern Observatory, Karl-Schwarzschild-Strasse, 2, 85748 Garching-bei-München,
Germany}
\email{chummel@eso.org}

\author[0000-0002-2763-7250]{Y. R. Cochetti} 
\affiliation{Departamento de Espectroscopía Estelar, Facultad de Ciencias Astronómicas y Geofísicas, Universidad Nacional de La Plata, La Plata, Argentina}
\affiliation{Instituto de Astrofísica de La Plata (CONICET-UNLP), Paseo del Bosque S/N, La Plata, Argentina}
\email{cochetti@fcaglp.unlp.edu.ar}

\author[0000-0002-3380-3307]{John D. Monnier} 
\affiliation{Astronomy Department, University of Michigan, Ann Arbor, MI 48109, USA}
\email{monnier@umich.edu}

\author[0000-0001-6017-8773]{Stefan Kraus} 
\affiliation{Astrophysics Group, Department of Physics \& Astronomy, University of Exeter, Stocker Road, Exeter, EX4 4QL, UK}
\email{S.Kraus@exeter.ac.uk}

\author[0000-0002-2208-6541]{Narsireddy Anugu} 
\affiliation{The CHARA Array of Georgia State University, Mount Wilson Observatory, Mount Wilson, CA 91023, USA}
\email{nanugu@gsu.edu}

\author[]{Sorabh Chhabra} 
\affiliation{Astrophysics Group, Department of Physics \& Astronomy, University of Exeter, Stocker Road, Exeter, EX4 4QL, UK}
\email{s.chhabra@exeter.ac.uk}

\author[0009-0005-8088-0718]{Isabelle Codron} 
\affiliation{Astrophysics Group, Department of Physics \& Astronomy, University of Exeter, Stocker Road, Exeter, EX4 4QL, UK}
\email{ic302@exeter.ac.uk}

\author[]{Claire L. Davies} 
\affiliation{Astrophysics Group, Department of Physics \& Astronomy, University of Exeter, Stocker Road, Exeter, EX4 4QL, UK}
\email{C.Davies3@exeter.ac.uk}

\author[0000-0002-3003-3183]{Tyler Gardner} 
\affiliation{Cooperative Institute for Research in Environmental Sciences, 216 UCB, University of Colorado Boulder,  Boulder, CO, 80309, USA}
\email{Tyler.Gardner@colorado.edu}

\author[0009-0006-0225-4444]{Mayra Gutierrez} 
\affiliation{Astronomy Department, University of Michigan, Ann Arbor, MI 48109, USA}
\email{mgutie60@ucsc.edu}

\author[]{Noura Ibrahim} 
\affiliation{Astronomy Department, University of Michigan, Ann Arbor, MI 48109, USA}
\email{inoura@umich.edu}

\begin{abstract}
    We study the nearby young ($\sim100$ Myr) moving group known as the Pisces-Eridanus (Psc-Eri) stellar stream, which contains the massive white dwarf KUV 0352+0500, and  $\tau^1$ Aqr, a bright double-lined spectroscopic and interferometric binary. $\tau^1$ Aqr has a B9V-type primary star nearing the group's main sequence turnoff. Binary modeling produces an orbital period $P = 28.07536 \pm 0.00001$ days, and masses $M_1 = 2.56^{+0.057}_{-0.052}$ $M_\odot$ and $M_2 = 1.28^{+0.069}_{-0.064}$ $M_\odot$. 
Additionally, we derive a distance $d = 103 \pm 1.1$ pc from the orbital parallax, indicating $\tau^1$ Aqr is closer than the Gaia DR3 distance estimate of $109.01$ pc. We also examine the spectral energy distributions (SEDs) of the bright members of Psc-Eri to derive temperatures and radii, and additionally constrain the stream's age. The age suggested by this analysis is $100-120$ Myrs, which is consistent with previous studies of the stream.
A preliminary analysis of white dwarf KUV 0352+0500 indicates an initial mass in the range of approximately $5.84-6.63$ $M_\odot$ and demonstrates that a more precise age would improve constraints on the high-mass end of the WD initial-final mass relation (IFMR).
\end{abstract}



\section{Introduction} \label{sec:intro}

A star's fate is determined by its initial mass, and it is well understood that low-mass stars evolve to produce white dwarfs, while high-mass stars end their lives in supernovae. The maximum mass of a white dwarf, known as the Chandrasekhar limit, is found from detailed models to be approximately $1.38$ $M_{\odot}$ for a carbon-oxygen composition \citep[e.g.][]{althaus}. However, the maximum mass of a white dwarf progenitor star is less well defined and remains a topic of active research \citep{Cummings2018,ElBadry,miller26}. The relationship between the initial mass of a progenitor star and the final mass of the resulting WD is called the WD initial-final mass relation (IFMR), where higher-mass single stars produce higher-mass WDs.
 
Young star clusters and moving groups provide the means to improve the definition of the high-mass end of the IFMR. These sometimes contain WDs that evolved from massive stars and are still relatively hot and bright, making them easier to study. A precise age for a group with a massive WD can produce a precise lifetime for the progenitor of the WD after subtracting its cooling time. Precise characterization of a massive binary star group member can contribute greatly to this effort by helping constrain the age and by providing a definitive measurement of mass for a massive, still evolving star. We studied the Pisces-Eridanus (Psc-Eri) stellar stream because it is a young group containing both a WD and a bright and detached binary star system that can be used for an age determination.

\subsection{Psc-Eri Group Identification and Age Review} \label{sec:psceri} 

Psc-Eri is a specific type of moving group called a stellar stream, and is likely the remnant of a star-forming region. This stream has a length up to 700 pc, with the nearest members as close as 80 pc and the furthest possibly up to 400 pc from the Sun \citep{roser}. As a result of these characteristics, the stream spans $120^{\circ}$ on the sky and is not obviously clustered. The stream was found to have a near-solar metallicity [Fe/H]$=-0.03$ \citep{metalpsceri}.

Psc-Eri was first identified by \cite{MGASTref} as a clustering in 3-D velocity space in a sample within 300 pc of the Sun in Gaia Data Release 2 (DR2). They subsequently used the DBSCAN clustering algorithm \citep{DBSCAN} to search for clusters in 6-D cylindrical phase space, and subsequent analysis resulted in a list of 256 members.
Initially, the age of Psc-Eri was reported as 1 Gyr by \cite{MGASTref}, though this age determination was heavily dependent on a G8IV sub-giant star, 42 Ceti. Soon after, \cite{MAINcurt} age dated the stellar stream using gyrochronology for
101 members observed by TESS, finding that the rotation period distribution overlaps with the distribution of the Pleiades. This implied that the stream is roughly 120 Myr old and that 42 Ceti was likely an interloper. A more recent study by \cite{roser} identified 1387 members of the stream and used isochrones to determine an age of 135 Myr, aligning with the younger age found by \cite{MAINcurt}. \cite{ratzen} found about 2000 members across the entire stellar mass spectrum, including white dwarfs.  
The Montreal Open Clusters and Associations (MOCA) database \citep{2026MOCA} lists 1126 plausible members (with 308 high likelihood candidate members and 818 candidate members) as of September 2026.

One probable WD member of Psc-Eri, with $<1\%$ probability of belonging to the background population and characteristics consistent with the age of the cluster is DA star KUV 03520+0500 \citep{roser}. This WD has had a few previous mass measurements: $1.12\pm0.02 M_\odot$ \citep{homeier}, $0.97\pm0.05 M_\odot$ \citep{vennes}, $1.09 \pm 0.03 M_{\odot}$ \citep{gianninas} and $1.10 M_\odot$ \citep{limoges} from independent ground-based Balmer-line spectra, and $1.201\pm0.016 M_\odot$ using much lower resolution Gaia DR3 XP spectra \citep{vincent}. The Balmer line studies agree on an effective temperature of around 37000 K, supporting its relative youth --- the only quoted cooling age for the WD is 60 Myr \citep{limoges}. There have not been indications of a strong magnetic field or unusual composition that might indicate that the WD resulted from a merger. Gaia parallax measurements put it at a distance of $111.6\pm0.7$ pc \citep{BailerJones}.

Our binary target in Psc-Eri is $\tau^1$ Aqr (HD 215766), which contains a B9V-type primary star. The system was identified as a double-lined spectroscopic binary by \citet{BESOspect}, but to date it has not been thoroughly studied and the nature of the secondary star was unknown.
$\tau^1$ Aqr falls close to the main sequence turnoff point of the group. This is where stellar masses start to become helpful for age determination due to the accelerating evolution changes in the star’s characteristics. For this reason, $\tau^1$ Aqr's bright primary star makes it a great candidate to use for age dating Psc-Eri.

\section{Data} \label{sec:data}

Mass measurements of the components of $\tau^1$ Aqr are an important ingredient for a more precise age of Psc-Eri. The masses we derive are from fits to a combination of spectroscopic radial velocity measurements and interferometric observations. 

\subsection{Radial Velocities from Spectroscopy} \label{sec:RVspec}

The radial velocities for $\tau^1$ Aqr were measured from archival spectra obtained using three different spectrographs. The first set of spectra was obtained between July 2009 and July 2012 using BESO \citep{BESOspect}, a high-resolution ($R = 48000$) spectrograph at the Hexapod Telescope (1.5 m) of Cerro Armazones Observatory (OCA). There were 11 observations in total, covering a wavelength range of $3526-8535$ \AA. X-SHOOTER\footnote{Based on observations made with ESO Telescopes at the La Silla Paranal Observatory under programme ID 60.A-9022 and obtained from the ESO Science Archive Facility} \citep{XSHOOTER}, a medium-resolution ($R = 4000-17000$) spectrograph mounted at the UT3 Cassegrain focus of the Kueyen telescope (8.2 m) was used between July and August 2010. Three spectra that had been processed with the standard pipeline xshoo/2.7.0b were collected from the ESO Archive. Each spectrum involved multiple wavelength sections, covering a wavelength range of $2989-10200$ \AA. Finally, the most recent observations from November 2015 to August 2016 (PI: Daniela Iglesias) used UVES\footnote{Based on observations made with ESO Telescopes at the La Silla Paranal Observatory under programme ID 097.C-0409 and obtained from the ESO Science Archive Facility}, a high-resolution [$R = 80000$ (blue arm), $R = 110000$ (red arm)] optical spectrograph of the VLT located at the Nasmyth B focus of UT2. There were a total of four observations, and the processed spectra (uves/5.10.0 pipeline) were also obtained from the ESO Archive. Similarly to the spectra from X-SHOOTER, three were observed in multiple sections and were later merged, covering the wavelength range of $3281-6835$ \AA.

To measure radial velocities, the processed spectra from each of these sources were continuum normalized, and analyzed using the program \texttt{BF-python}\footnote{\texttt{BF-python} GitHub page: https://github.com/mrawls/BF-rvplotter \citep{bfpython} that computes the broadening function $f$ \citep{Ruc92,Ruc99}}. The broadening function is a convolution function that satisfies $B = f \otimes N$, where $B$ is the observed spectrum, and $N$ is a narrow-lined spectrum template, which can either be a high-resolution observed spectrum or theoretical spectrum of a star with similar effective temperature ($T_{\rm eff}$), metallicity, and gravity. This function can reveal the radial velocity of the stars, as well as their rotational broadening.

\begin{figure}[ht!] 
\centering
    \includegraphics[width=1\linewidth]{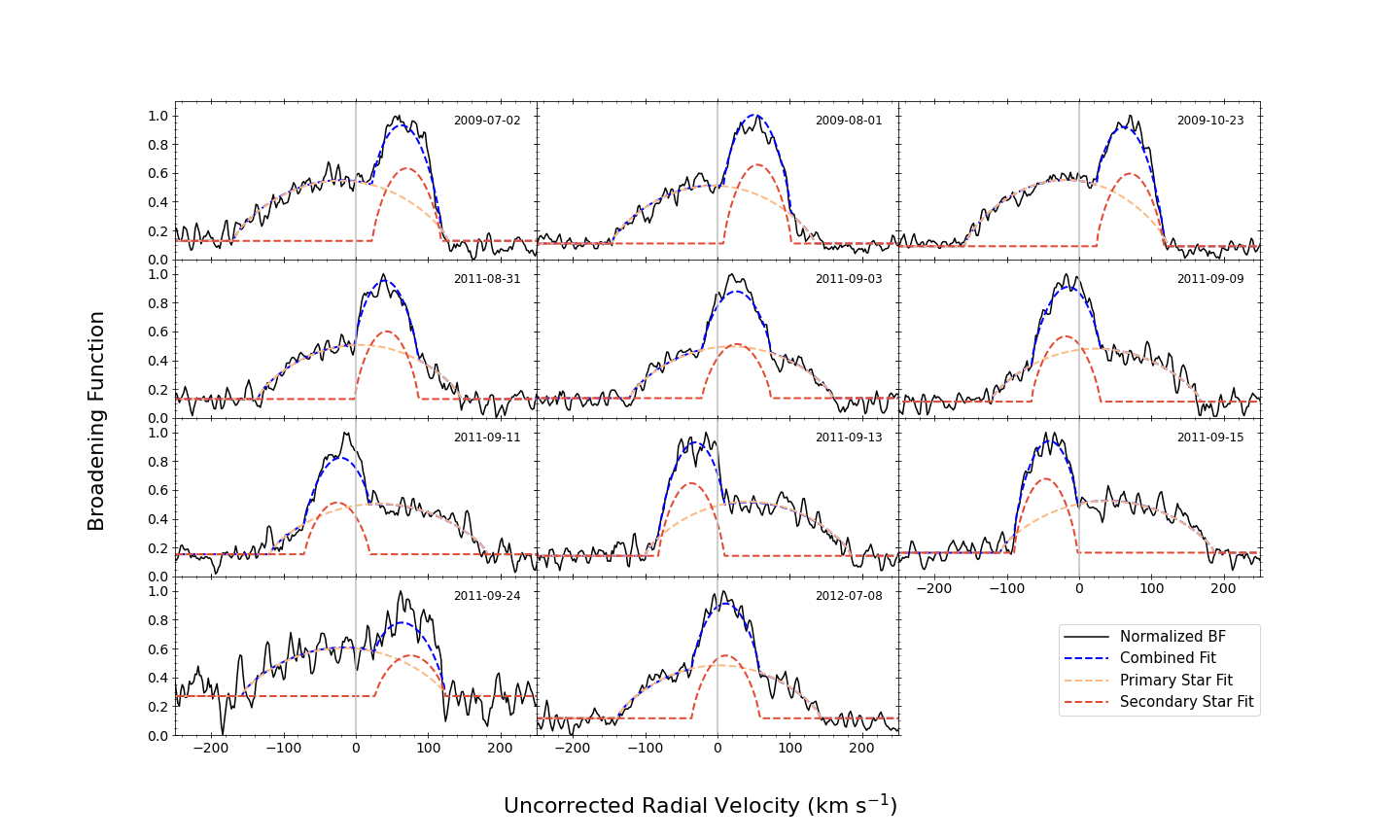}
\caption{Normalized broadening functions for the BESO spectra.} 
\label{fig:BESOBF}
\end{figure}

We used a synthetic spectrum template from the BLUERED library \citep{synthspect_temp} in our analysis. The computed broadening function was fitted with theoretical rotational broadening function profiles in order to measure the rotational velocities of both components\footnote{Rotational broadening function fitting capability was added to \texttt{BF-python} by the authors.}. Figure \ref{fig:BESOBF} displays the broadening functions for the BESO spectra, where the brighter primary star is the broader and faster-rotating component, and the secondary star is a slower rotator with narrower lines. Due to the distinct differences in rotational broadening, the radial velocities of the two stars could be reliably measured at any orbital phase.

In the process of measuring radial velocities, some discrepancies were found in the wavelength calibrations of the BESO spectra. We cross-correlated approximately 30 absorption lines in an atmospheric O$_2$ molecular band ($6270-6320$ \AA) for each of the spectra against those in the spectrum with the least noise in order to find radial velocity shift corrections. Before combining the radial velocity datasets from the different spectrographs, we fitted the radial velocities from each spectrograph individually with Keplerian orbits to determine systematic velocity offsets. Offsets were applied to the UVES and X-SHOOTER sets to put them on the same system as the largest BESO dataset. 

The measured radial velocities are listed in Table \ref{tab:RVs}, along with five additional primary star radial velocities measured between September 1912 and November 1949 at Mount Wilson Observatory (MWO) found in the literature \citep{MWORV}. The MWO observations were excluded from the orbital modeling because they only provided radial velocity measurements for the primary star and likely contained systematic errors due to the undetected secondary star. We determined the rotational velocities of both components by taking the average value measured specifically from the observations where the edges of both component's profiles are most clearly defined. We find a rotational velocity of $143$ km/s for the primary and $46$ km/s for the secondary. 

\begin{deluxetable}{lccccl}
\tablewidth{0pc}
\tablecaption{Radial Velocities for $\tau^1$ Aqr \label{tab:RVs}}
    \tablehead{ \colhead{mJD\tablenotemark{a}} & \colhead{$v_{r,1}$} &  \colhead{$\sigma_1$} & \colhead{$v_{r,2}$} &  \colhead{$\sigma_2$} & \colhead{Source} \\
    & \multicolumn{2}{c}{(km s$^{-1}$)} & \multicolumn{2}{c}{(km s$^{-1}$)} &}
\startdata
        19647.373 & 10.6 & 4.1\tablenotemark{b} & - & - & MWO  \\
        32844.184 & 12.2 & 2.0\tablenotemark{b} & - & - & MWO  \\
        33164.311 & 23.9 & 2.0\tablenotemark{b}	& - & - & MWO  \\
        33167.250 & 28.2 & 0.8\tablenotemark{b} & - & - & MWO  \\
        33231.132 & 1.6	& 1.9\tablenotemark{b} & - & - & MWO   \\ 
        \hline
        55014.92824 & $-23.71$ & 4.07 & 67.63    & 1.65 & BESO     \\
        55044.81375 & $-9.25$  & 3.26 & 51.94    & 1.24 & BESO     \\
        55127.63550 & $-21.91$ & 3.45 & 66.50    & 1.52 & BESO     \\
        55804.23787 & $-0.10$  & 3.32 & 36.67    & 1.43 & BESO     \\
        55807.71323 & 7.37     & 3.83 & 14.87    & 1.49 & BESO     \\
        55813.68258 & 21.21    & 4.28 & $-20.34$ & 1.40 & BESO     \\
        55815.67635 & 23.52    & 4.54 & $-31.23$ & 2.02 & BESO     \\
        55817.66139 & 38.43    & 4.12 & $-39.92$ & 1.79 & BESO     \\
        55819.66039 & 38.27    & 4.03 & $-46.61$ & 1.74 & BESO     \\
        55828.66328 & $-21.36$ & 6.22 & 70.73    & 2.99 & BESO     \\
        56116.82515 & 2.25     & 3.78 & 11.52    & 1.67 & BESO     \\ 
        \hline
        55408.79834 & $-11.04$ & 3.38 & 59.69    & 1.56 & X-SHOOTER \\
        55410.83662 & $-5.12$  & 3.01 & 39.66    & 1.95 & X-SHOOTER \\
        55410.83762 & $-2.40$  & 3.20 & 36.76    & 1.94 & X-SHOOTER \\ 
        \hline
        57340.57246 & 1.75     & 9.09 & 52.18    & 0.37 & UVES     \\
        57585.91877 & 38.97    & 4.66 & $-37.75$ & 0.59 & UVES     \\
        57595.89857 & $-36.63$ & 3.99 & 95.02    & 0.36 & UVES     \\
        57611.86024 & 28.01    & 4.38 & $-28.18$ & 0.63 & UVES     \\
\enddata
\tablenotetext{a}{mJD = HJD $- 2400000$}
\tablenotetext{b}{MWO uncertainties are not scaled and are likely underestimates.}
\end{deluxetable}

\subsection{Interferometric Data} \label{sec:interfdat}

Interferometric observations over time can trace the relative plane-of-the-sky motions of stars in resolved binaries. There are four nights of archived observations taken with the four-telescope VLTI-PIONIER interferometric combiner \citep{pionierpap} in $H$ band with baselines ranging from 22 to 317 m. One observation from November 2012 recorded data in three spectral channels (PI: M. Rabus)\footnote{Based on observations collected at the European Southern Observatory under ESO programme 090.D-0917(C)} and three observations from July to September 2017 recorded data in 6 channels (PI: M. Schoeller)\footnote{Based on observations collected at the European Southern Observatory under ESO programme 099.D-0513(B)}. $\tau^1$ Aqr was used as a calibration target in the original observations, but they were recalibrated against other unresolved targets measured on the same nights.

We obtained four additional observations using the six-telescope CHARA Array \citep{tenBrummelaar} with the MIRC-X \citep[$H$ band;][]{Anugu} and MYSTIC \citep[$K$ band;][]{Setterholm} beam combiners, with 15 baseline combinations ranging from $34 - 331$ m in length. Three were taken in September 2023 and one in September 2024 using the Prism $R=50$ and Prism $R=49$ dispersion elements, respectively for the two combiners, providing eight spectral channels per beam combination. Information about the interferometric observations can be found in Table \ref{tab:interfdata}.




\begin{deluxetable}{ccccccl}
\tablewidth{0pc}
\tablecaption{Interferometric Observations of $\tau^1$ Aqr \label{tab:interfdata}}
    \tablehead{\colhead{UT Date} & \colhead{mJD\tablenotemark{a}}  &  \colhead{Combiner} & \colhead{$N_{tel}$} &  \colhead{$N_{vis}$} & \colhead{$N_{phase}$} & \colhead{Calibrators\tablenotemark{b}}}
    \startdata
      2012 Nov 23   & 56254.51794 &  VLTI-PIONIER & 4 & 54   & 36   & BD$-15$ 6290\tablenotemark{c} \\ \hline
      2017 July 21  & 57955.87000 &  VLTI-PIONIER & 4 & 72   & 48   & HD 31998,\\ 
                     &            &               &   &      &      & HD 31721 \\ \hline
      2017 Aug 30   & 57995.75500 &  VLTI-PIONIER & 4 & 72   & 48   & HD 215782\\ \hline
      2017 Sept 2   & 57998.80000 &  VLTI-PIONIER & 4 & 72   & 48   & HD 30734 \\ \hline
      2023 Sept 14  & 60201.69169 &  CHARA-MIRCX   & 6 & 1920 & 2400 & HD 206893,  \\
       &  &  CHARA-MYSTIC  & 6 & 2309 & 2788 & HD 223448 \\ \hline
      2023 Sept 18  & 60205.72421 &  CHARA-MIRCX   & 6 & 512  & 448  & HD 206893, \\ 
       &   &  CHARA-MYSTIC  & 6 & 640  & 560  & HD 215464,\\
       &  &   &   &  &  & HD 223448\\ \hline
      2023 Sept 28  & 60215.67976 &  CHARA-MIRCX   & 6 & 1200 & 2240 & HD 206893 \\
       &  &  CHARA-MYSTIC  & 6 & 1005 & 1330 & HD 223448 \\ \hline
      2024 Sept. 6 & 60559.69112 & CHARA-MIRCX & 6 & 444 & 579 & HD 206893\\
                   &             & CHARA-MYSTIC & 6 & 563 & 736 & HD 223448\\
    \enddata
\tablenotetext{a}{mJD = HJD$- 2400000$}
\tablenotetext{b}{Angular diameters ($\theta_\mathrm{LDD}$) of the calibration stars from Version 2 of the JMMC Stellar Diameters Catalogue \citep{Bourges2017}: 
HD 206893 ($0.3075 \pm 0.0076$ mas); HD 215464 ($0.1880 \pm 0.0049$ mas); HD 223448 ($ 0.3462 \pm 0.0078$ mas). From the JMMC Measured Stellar Diameters Catalog \citep{duvert}: BD$-15$ 6290 ($0.705 \pm 0.0009$ mas).}
\tablenotetext{c}{ BD$-15$ 6290 was the original science target, but was later used as a calibrator for $\tau^1$ Aqr.}
\end{deluxetable}

The observables from the interferometric observations were fringe visibilities and closure phases. The visibility of a uniformly bright disk is 
\begin{equation}
    V = \frac{2 J_1 (\pi B \theta/ \lambda)} { \pi B \theta / \lambda}
\end{equation}
where $J_1$ is the first-order Bessel function, the baseline $B$ is the distance between pairs of telescopes, $\theta$ is the angular diameter of the star, and $\lambda$ is the wavelength of the observed light. The visibility for a binary star is more complex, as the pattern is produced by the overlapping interference patterns of light from both stars instead of one. The squared visibility $V^2$ is
\begin{equation}
    V^2 = \frac{V_1^2 + r^2 V_2^2 + 2r|V_1||V_2| \cos{[2 \pi \boldsymbol{B} \cdot \frac{\boldsymbol{s}}{\lambda}]}}{(1 + r)^2}
\end{equation}
in which $V_1$ and $V_2$ are the visibilities of the two stellar components, $r$ is the brightness ratio, $\boldsymbol{B}$ is the baseline vector, and $\boldsymbol{s}$ is the angular separation vector \citep{Lawson}. The measured visibility along each baseline gives the projected separation along the baseline vector as well as the flux ratio. Measurements of visibilities along multiple baselines yields the separation and position angle in two dimensions as projected on the sky, along with multiple measurements of the flux ratio.

Closure phases are another useful observable when studying binary systems. Phase in interferometry refers to the position of the peak of the interference fringes within the broader envelope. This envelope is a Gaussian-like curve that surrounds the smaller interference fringes, caused by the finite passband of the interferometer. The phase tells us where the peak of one of these fringes falls within the envelope, and because interferometry uses multiple telescopes, we can measure the difference in phase between pairs of telescopes.  These phases ($\phi$) provide information about the object’s orientation relative to the baseline. Unfortunately, the phase from a single baseline is affected by atmospheric effects.
A closure phase ($\Phi$) is calculated by combining the phases from baselines for a set of three telescopes observing the same target. This combination is
\begin{equation}
    \Phi_{1,2,3} = \phi_{1,2} + \phi_{2,3} + \phi_{3,1}
\end{equation}
where $\phi_{1,2}$ is the phase from the baseline between telescopes 1 and 2. The benefit of closure phases is that they cancel the systematics caused by atmospheric turbulence above any single telescope, leaving more robust information about the target. Closure phases are highly sensitive to asymmetries in the light distribution. Point-symmetric light distributions, such as single stars or binary systems with components of identical brightness, yield closure phases of 0 or 180 degrees. In contrast, binary systems with unequal components create an asymmetrical distribution, resulting in non-zero (and non-180) closure phases that are highly sensitive to the sky orientation and brightness ratio. This information revealed by both observables (visibility amplitudes and closure phases) provides strong geometrical insights into the orbital configuration of the system. This is vital for determining a binary's inclination, a key parameter in calculating the masses of both components of $\tau^1$ Aqr.

The MIRC-X and MYSTIC data were reduced using the standard pipeline\footnote{\url{https://gitlab.chara.gsu.edu/lebouquj/mircx_pipeline}} (version 1.4.0 and 1.5.0; \citealt{Anugu}) using two different integration times of 30 s and 150 s. The 30 s data were calibrated using the IDL script written by J. D. Monnier; the deep cleaning option was used to remove outliers, and the data were averaged over 6 min intervals. The 150 s integrations were calibrated using the MIRC-X/MYSTIC calibration script written in python. The uniform disk angular diameters in the $H-$ and $K-$bands for the calibrators were adopted from the Jean-Marie Mariotti Centre (JMMC) Stellar Diameter Catalogue (JSDC) Version 2 \citep{Bourges,Chelli}. The pipeline yields squared visibilities and closure phases for all baselines and spectral channels. For both MIRC-X and MYSTIC observations, a 5\% uncertainty on visibility measurements and a 0.3 deg uncertainty on closure phases were added in quadrature with formal errors to account for calibration uncertainties. During the fitting process, the wavelengths in the OIFITS files were divided by systematic correction factors \citep[e.g.,][]{gardner_mirc} of 1.0054 $\pm$ 0.002 for MIRC-X and 1.0067 $\pm$ 0.002 for MYSTIC (J.D. Monnier, priv. comm.), which affects the inferred spatial scale for  position measurements. 
PIONIER data reduction used the pndrs pipeline \citep{pionierpap}. The 2012 data were obtained from the JMMC Optical Interferometry Database \citep[OiDB][]{oidb}, already processed through the standard pipeline. This pipeline used an average (weighted inversely by separation in time) of the transfer functions for calibrator stars during the night, but the closest calibrator is listed in Table \ref{tab:interfdata}. 
A 1.5\% systematic uncertainty in the effective wavelength of the observations contributes significantly to the uncertainty in the relative positions.

\begin{deluxetable}{ccccccccc}
\tablewidth{0pc}
\tablecaption{Interferometric Position Measurements of $\tau^1$ Aqr \label{tab:position}}
    \tablehead{ \colhead{mJD\tablenotemark{a}} & \colhead{$\rho$ (mas)} & \colhead{$\theta$ (deg)} & \colhead{$\Delta \alpha$ (mas)} & \colhead{$\Delta \delta$ (mas)} & \colhead{$\sigma_{maj}$} & \colhead{$\sigma_{min}$} & \colhead{$\varphi$} & \colhead{Combiner}}
\startdata
56254.548 & 1.859 & 23.28 & 0.735    & 1.708    & 0.038 & 0.033 & 0.00   & PIONIER \\
57955.870 & 1.840 & 130.60 & 1.397    & $-1.198$ & 0.025 & 0.025 & 0.00   & PIONIER \\
57995.755 & 1.880 & 32.50 & 1.010    & 1.586    & 0.025 & 0.025 & 0.00   & PIONIER \\
57998.800 & 2.330 & 59.50 & 2.008    & 1.183    & 0.025 & 0.025 & 0.00   & PIONIER \\
60201.697 & 1.892 & 128.68 & 1.477    & $-1.182$ & 0.015 & 0.006 & 141.37 & MIRC-X   \\
60201.697 & 1.899 & 128.67 & 1.482    & $-1.187$ & 0.024 & 0.009 & 131.12 & MYSTIC  \\
60201.742 & 1.868 & 129.23 & 1.447    & $-1.181$ & 0.011 & 0.006 & 137.97 & MIRC-X   \\
60201.743 & 1.871 & 129.18 & 1.450    & $-1.182$ & 0.015 & 0.006 & 140.96 & MYSTIC  \\
60201.781 & 1.846 & 129.70 & 1.420    & $-1.180$ & 0.013 & 0.007 & 6.32   & MIRC-X   \\
60201.781 & 1.850 & 129.77 & 1.422    & $-1.183$ & 0.013 & 0.011 & 25.52  & MYSTIC  \\
60205.729 & 1.505 & 296.99 & $-1.341$ & 0.683    & 0.092 & 0.025 & 135.61 & MIRC-X   \\
60205.729 & 1.556 & 297.49 & $-1.380$ & 0.718    & 0.067 & 0.019 & 137.53 & MYSTIC  \\
60205.779 & 1.620 & 298.25 & $-1.427$ & 0.767    & 0.092 & 0.022 & 115.26 & MIRC-X   \\
60205.779 & 1.568 & 298.18 & $-1.382$ & 0.740   & 0.121 & 0.024 & 112.01 & MYSTIC  \\
60215.686 & 2.114 & 51.27 & 1.650    & 1.323    & 0.012 & 0.007 & 167.11 & MIRC-X   \\
60215.687 & 2.117 & 51.33 & 1.653    & 1.323    & 0.019 & 0.009 & 155.97 & MYSTIC   \\
60215.730 & 2.118 & 51.72 & 1.662    & 1.312    & 0.013 & 0.005 & 154.95 & MIRC-X   \\
60215.730 & 2.134 & 51.29 & 1.665    & 1.335    & 0.027 & 0.007 & 164.27 & MYSTIC   \\
60559.695 & 3.054 & 92.67 & 3.051    & $-0.142$ & 0.007 & 0.003 & 141.870 & MIRC-X   \\
60559.695 & 3.071 & 92.89 & 3.067    & $-0.155$ & 0.009 & 0.002 & 139.100 & MYSTIC   \\
60559.695 & 3.056 & 92.61 & 3.053    & $-0.139$ & 0.007 & 0.003 & 141.826 & MIRC-X   \\
60559.695 & 3.067 & 92.86 & 3.063    & $-0.153$ & 0.008 & 0.002 & 138.954 & MYSTIC  \\
\enddata
\tablenotetext{a}{mJD = HJD$- 2400000$}
\end{deluxetable} 

The CHARA positions were obtained using a grid search\footnote{For more information, \url{https://www.chara.gsu.edu/analysis-software/binary-grid-search} \citep{chara_gridsearch}} that fits binary models to the visibilities, closure phases, and the $uv$ coordinates (which refers to the projected separation and direction of each baseline with respect to the target on the sky) when the measurements were made \citep{binarygrid}. The code varies the separations in right ascension ($\Delta \alpha$) and declination ($\Delta \delta$) and optimizes the binary position and flux ratio at each step in the grid. The solution with the minimum $\chi^2$ is adopted as the best fit.
Positions for the VLTI-PIONIER interferometric data were obtained using the software \texttt{LITpro} \citep{LITpropaper}, which follows a similar grid search process.
The resulting position measurements are given in Table \ref{tab:position} by angular separation ($\rho$) and position angle ($\theta$), as well as right ascension ($\alpha$) and declination ($\delta$) offsets of the secondary star from the primary. We also provide the major ($\sigma_{maj}$) and minor ($\sigma_{min}$) axis uncertainties and the rotation ($\varphi$) of the error ellipses. 

\section{Analysis} \label{sec:analysis}
\subsection{Orbital Modeling of $\tau^1$ Aqr} \label{sec:orbmod}

We used a combination of radial velocity data and geometric information from interferometric measurements of $\tau^1$ Aqr in modeling the binary system and its orbit. We use a total of 12 parameters in the orbital modeling: binary period ($P$), reference time for periastron ($t_0$), radial velocity semi-amplitudes of the stars ($K_1, K_2$), binary system velocity ($\gamma$), orbit eccentricity ($e$), argument of periastron ($\omega$), position angle of the ascending node ($\Omega$), orbit inclination ($i$), angular size of the semi-major axis on sky ($a^{\prime\prime}$), and the stellar luminosity ratios in $H$ and $K$ bands $r_H = L_2 (H) / L_1 (H)$ and $r_K = L_2 (K) / L_1 (K)$. The values of these parameters are sampled using \texttt{interfRVorbit}, which uses a genetic search algorithm \citep{genetic_alg}. 

The radial velocities and interferometric data (visibilities and closure phases) or relative positions were modeled with \texttt{interfRVorbit} in different combinations. We present three fits: the first using only RV data; the second fitting the RV data and the interferometric observables from all epochs simultaneously; and the third fitting the RVs and the individual-epoch astrometry derived previously in Section \ref{sec:interfdat}. Modeling the orbit with these three sets of data allows for cross-verification of the results and checks on the effects of the weighting (by measurement uncertainties) of the different data types.

Properly weighting each dataset is important because of the large differences in numbers of observations and the differing constraints each type of data puts on the orbital parameters. To account for this, we applied a scaling factor to the uncertainties associated with each data type (RVs, visibilities, closure phases, and positions) to ensure the reduced $\chi^2$ value ($\chi_\nu^2$) was approximately 1 for a fit to that data alone. 
Once a satisfactory final fit to the data with scaled uncertainties was obtained, we determined the orbital parameter uncertainties from the ranges covered by trial models having $\chi^2$ within one of the minimum $\chi^2$ \citep{1sig_error}.

Seven parameters can be constrained by radial velocity measurements alone: period $P$, reference time of periastron $t_0$, velocity semi-amplitudes of the primary $K_1$ and secondary $K_2$, eccentricity $e$, argument of periastron $\omega$, and binary system velocity $\gamma$. Our initial fits involved accounting for velocity offsets between the datasets from different spectrographs. RV uncertainties for BESO measurements were scaled based on their scatter around the best fitting model. RV uncertainties from X-SHOOTER and UVES could not be scaled on their own because of the small number of measurements available and their small span in orbital phase, and so they were instead scaled the same as the BESO uncertainties. 
Modeling the orbit with the radial velocities alone allowed us to check the scaling of the uncertainties and refine the ephemeris of the system. This fit resulted in a $\chi_\nu^2 = 1.22$, and the fitted parameters are given in Table \ref{tab:orbparams}.

The parameters derived from the RV-only fit were used as initial guesses for the fit to the radial velocities plus interferometric visibilities and closure phases. This combined dataset allows for a fit to all 12 parameters. For this fit, the fixed angular diameters we assume are $\theta_1 = 0.191$ mas for the primary, obtained from SED fitting of the primary star photometry discussed in Section \ref{brightstars}, and $\theta_2 = 0.107$ mas for the secondary, estimated from isochrone radii based on the star's mass and distance from previous fits. The angular diameters only mildly influence the modeling of the interferometric data because the star diameters are unresolved across our interferometric baselines.\footnote{ A subsequent refinement of our SED fit presented in Section \ref{brightstars} yields the value $\theta_1 = 0.175$ mas for the primary, and using the secondary mass and distance from our final model results in $\theta_2 = 0.110$ mas. We ran the fit with these values and find these minor differences have a negligible impact on the orbital parameters, and do not change the results of our age analysis.}

To minimize systematic errors, the data from CHARA and VLTI were first modeled separately with the previously reweighted radial velocities to scale their respective uncertainties. These fits yielded $\chi_\nu^2 = 1.02$ for the VLTI-PIONIER data and $\chi_\nu^2 = 1.00$ for the CHARA data. Finally, we modeled the reweighted radial velocities with the reweighted interferometric data from both sources, producing a total $\chi_\nu^2 = 1.03$ with the fit parameters listed in Table \ref{tab:orbparams}.

The fit to the RV and position data was executed to check for consistency. The RVs and relative sky position data can be fit to ten of the 12 parameters (the brightness ratios are unconstrained). As with the fit to the RV and interferometric data, the VLTI-PIONIER and CHARA positions were separately modeled with the radial velocities to ensure the uncertainties were properly adjusted, achieving $\chi_\nu^2 = 1.62$ for VLTI-PIONIER and $\chi_\nu^2 = 1.00$ for CHARA. The final fit with both data sources resulted in a $\chi_\nu^2$ near one (in this case, 1.59), indicating a reliable fit. The resulting parameters agree well with the parameters from the other models, and are listed in Table \ref{tab:orbparams}.

\begin{deluxetable}{lcccccc}
\tablewidth{0pc}
\tablecaption{ $\tau^1$ Aqr Orbital Parameters \label{tab:orbparams}}
    \tablehead{ \colhead{Parameter} & \multicolumn{6}{c}{Fit} \\ & \multicolumn{2}{c}{RVs Only} & \multicolumn{2}{c}{RVs + Interferometry} &  \multicolumn{2}{c}{RVs + Positions}}
    \startdata
    $P$ (d)    & $28.0736$ & $^{+0.0007}_{-0.0004}$ & $28.07536$ & $^{+0.00001}_{-0.00001}$ & $28.07472$ & $^{+0.00007}_{-0.00006}$\\
    $t_0 - 2460204$ (HJD)   & $0.193$ & $^{+0.077}_{-0.041}$ & $0.26280$ & $^{+0.0002}_{-0.0004}$ & $0.274$ & $^{+0.0057}_{-0.0038}$\\
    $K_1$ (km/s)    &$38.67$ & $^{+1.31}_{-1.61}$ & $38.19$ & $^{+1.20}_{-1.16}$ & $38.22$ & $^{+1.49}_{-1.43}$\\
    $K_2$ (km/s)    & $77.73$ & $^{+0.27}_{-1.04}$ & $76.08$ & $^{+0.12}_{-0.18}$ & $76.51$ & $^{+0.27}_{-0.26}$\\
    $e$          & $0.558$ & $^{+0.002}_{-0.008}$ & $0.54197$ & $^{+0.00006}_{-0.00002}$ & $0.54227$ & $^{+0.0011}_{-0.0008}$\\
    $\omega$ ($^\circ$)    & $114.64$ & $^{+1.19}_{-0.44}$ & $113.574$ & $^{+0.004}_{-0.004}$ & $113.66$ & $^{+0.06}_{-0.08}$\\
    $\gamma$ (km/s)     & $9.5$ & $^{+0.3}_{-0.2}$ & $9.30$ & $^{+0.15}_{-0.003}$ & $9.93$ & $^{+0.20}_{-0.21}$\\
    $i$ ($^\circ$)     & - & - & $61.106$ & $^{+0.004}_{-0.0001}$ & $60.94$ & $^{+0.06}_{-0.07}$ \\
    $\Omega$ ($^\circ$)     & - & - & $118.419$ & $^{+0.0002}_{-0.000}$ & $118.46$ & $^{+0.08}_{-0.06}$\\
    $a^{\prime\prime}$ (mas)\tablenotemark{a}       & - & - & $2.7389$ & $^{+0.0055}_{-0.0055}$ & $2.732$ & $^{+0.007}_{-0.007}$\\
    $L_2 / L_1 (H)$     & - & - & $0.3559$ & $^{+0.0002}_{-0.000003}$ & - & -\\
    $L_2 / L_1 (K)$     & - & - & $0.3719$ & $^{+0.0004}_{-0.00002}$ & - & -\\
    $\theta_1$  (mas)   & - & - & $0.191$ &  (fixed)  & - & -\\ 
    $\theta_2$  (mas)   & - & - & $0.107$ &  (fixed)  & - & -\\
    $M_1 \sin^3 i$     & $1.750$ & $^{+0.042}_{-0.055}$ & $1.715$ & $^{+0.035}_{-0.036}$ &  $1.737$ & $^{+0.046}_{-0.044}$ \\ 
    $M_2 \sin^3 i$     & $0.871$ & $^{+0.050}_{-0.058}$ & $0.861$ & $^{+0.043}_{-0.046}$ & $0.868$ & $^{+0.057}_{-0.053}$ \\ 
    $M_1$ ($M_\odot$)    & - & - & $2.56$ & $^{+0.057}_{-0.052}$ & - & - \\
    $M_2$ ($M_\odot$)     & - & - & $1.28$ & $^{+0.069}_{-0.064}$ & - & -\\ 
    $M_{tot}$ ($M_\odot$)     & - & - & $3.84$ & $^{+0.090}_{-0.083}$ & $3.90$ & $^{+0.315}_{-0.370}$ \\ 
    \hline 
    $\Delta \gamma_{BESO-UVES}$ (km/s)& $-2.99$ & $\pm0.07$     &  &  \\ 
    $\Delta \gamma_{BESO-XSHOOTER}$ (km/s)& $-2.20$ & $\pm0.05$     &  &  \\ 
    \enddata
\tablenotetext{a}{A 0.2\% systematic wavelength calibration uncertainty was added in quadrature.}
\end{deluxetable}

\begin{figure}[ht] 
\centering
    \includegraphics[width=1\linewidth]{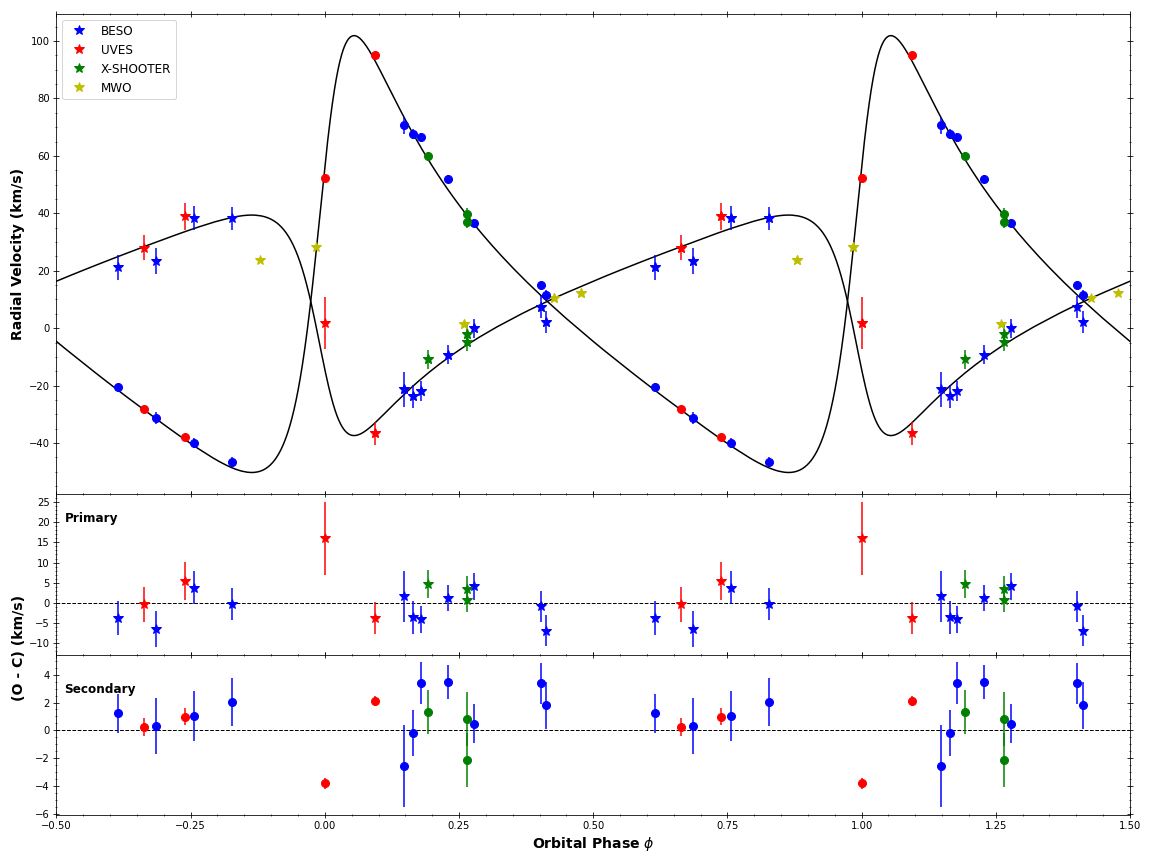}
\caption{\textit{Top panel:} Phased radial velocities and model curves from the RV and interferometric data fit. \textit{Stars} represent the primary RVs and \textit{dots} represent the secondary RVs. The MWO RVs are only plotted for reference, as they were not used in the orbital modeling. \textit{Middle/Bottom panels:} Radial velocity residuals (observed minus computed) for the primary and secondary stars.} 
\label{fig:RVcurve}
\end{figure}

\begin{figure}[p] 
\centering
    \begin{minipage}{\textwidth}
        \includegraphics[width=0.95\linewidth]{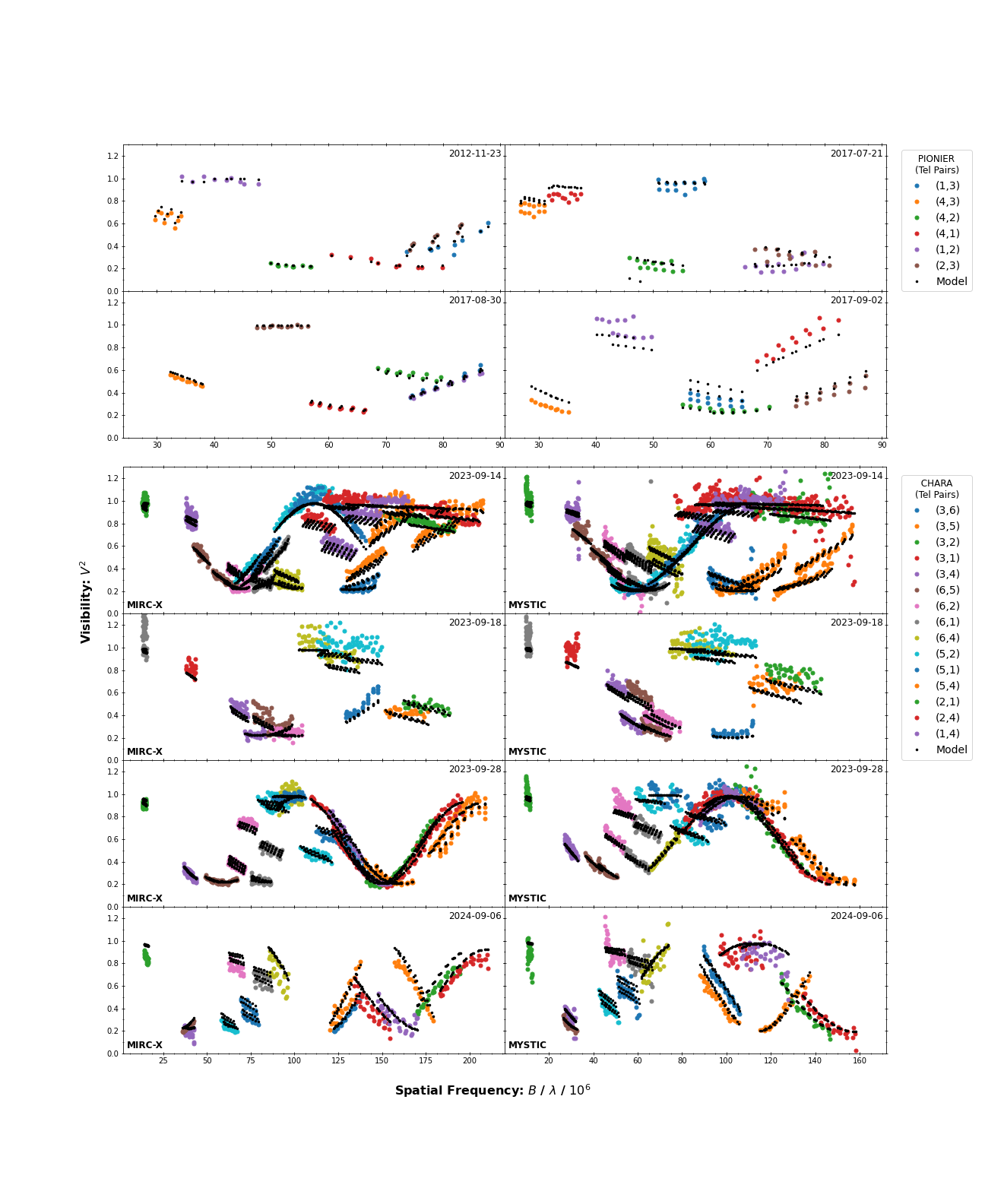}
        \caption{Observed and model predicted squared visibilities plotted against spatial frequency. The top four plots are from observations with PIONIER. The bottom six used CHARA, with MIRC-X in the left column and MYSTIC in the right. 
        } 
        \label{fig:vis}
    \end{minipage}
\end{figure}

\begin{figure}[p] 
\centering
    \begin{minipage}{\textwidth}
        \includegraphics[width=0.95\linewidth]{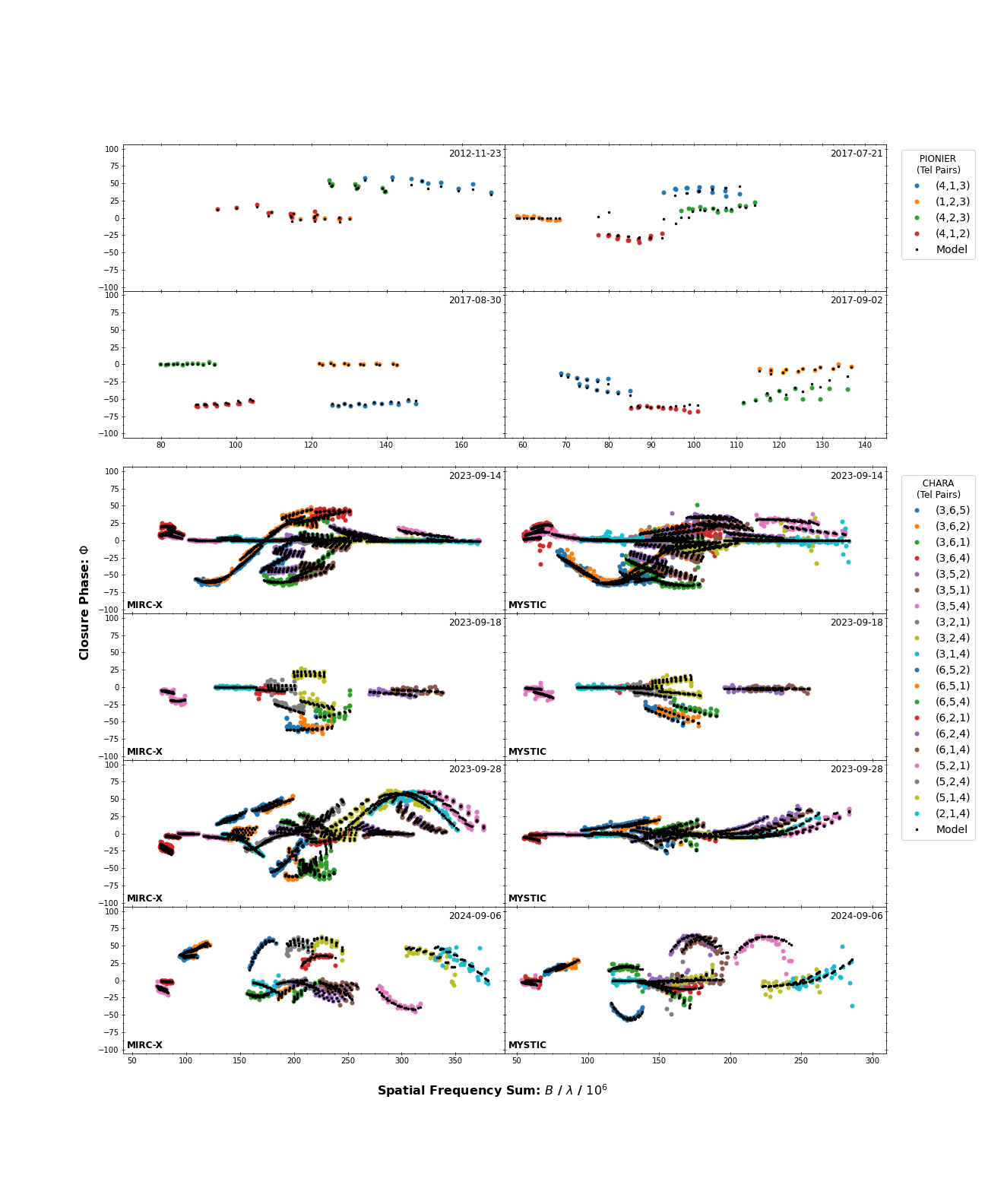}
        \caption{Observed and model predicted closure phases plotted against spatial frequency. The top four plots are from observations with PIONIER. The bottom six used CHARA, with MIRC-X in the left column and MYSTIC in the right.} 
        \label{fig:closphi}
    \end{minipage}
\end{figure}

\begin{figure}[ht!] 
\centering
    \includegraphics[width=1\linewidth]{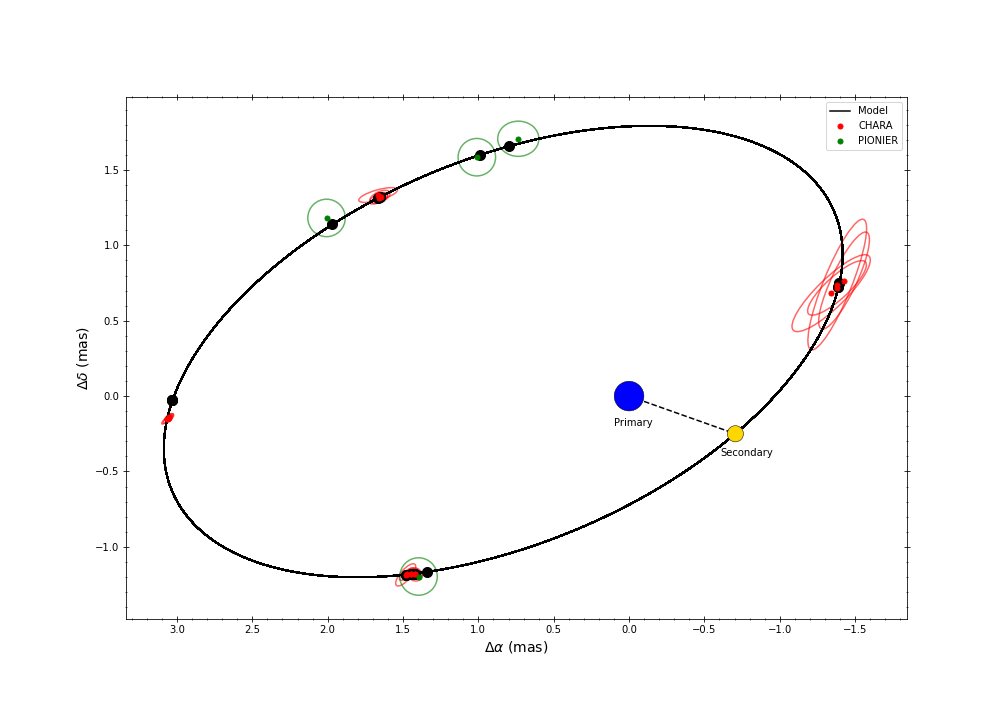}
\caption{Model of the relative orbit and positions of $\tau^1$ Aqr from the RV and interferometric data fit. The measured positions are shown with error ellipses expanded by a factor of 5 for visual clarity. The radii of the binary components are drawn to scale, and the secondary star is shown at the periastron position.} 
\label{fig:orbit}
\end{figure}

We compare the observed data and our orbital model fit of the RVs and interferometric observables in Figures \ref{fig:RVcurve} - \ref{fig:orbit}. As shown in the RV curve in Figure \ref{fig:RVcurve}, the model agrees well with the RVs, including the primary star measurements from MWO that were not used in the modeling process. Figures \ref{fig:vis} and \ref{fig:closphi} show the squared visibilities and closure phases plotted against their respective baselines for different dates, and the model fit values for comparison. Both plots show agreement across different baselines, suggesting that the model accurately represents the system's geometry and dynamics. Finally, Figure \ref{fig:orbit} shows the best-fit orbit of the secondary star with respect to the primary. The observed positions clearly agree with model predicted positions at the times of the observations. 

The radial velocity curve and orbital position plots (Figs. \ref{fig:RVcurve} and \ref{fig:orbit}) highlight a few limitations. The measurement of the RV semiamplitudes (and thereby, the masses) could be improved in the future with more observations near the extrema of the velocity curves. Additional interferometric observations near periastron would also more precisely constrain the semi-major axis and inclination of the orbit. 

The orbital solution provides an independent method for estimating the distance of the binary system. The orbital parallax is determined by comparing the angular semi-major axis $(a^{\prime \prime})$ to the physical size of the orbit, calculated from the period $(P)$, eccentricity $(e)$, inclination $(i)$, and velocity semi-amplitudes $(K_{1})$ and $(K_{2})$. We derive a distance $d = 103 \pm 1.1$  pc based on the resulting orbital parallax from our best-fit model. This value is less than the Gaia DR3 distance of $109.01$ pc \citep{BailerJones}. Larger velocity semi-amplitudes would point to a larger distance to match the measured angular size. Although our RV dataset lacks direct measurements at the velocity curve extrema, the significant increase to the semi-amplitudes necessary to match the Gaia distance would cause disagreement with the surrounding data points. Gaia parallaxes for partially resolved binary systems can be affected by their orbital motion. The current Renormalized Unit Weight Error (RUWE) for this system is 0.727 \citep{gaiadr3}. A value of RUWE $<1$ suggests an overestimation of the astrometric error \citep{Guerriero}, but more information is needed to interpret how this may be related to the distance discrepancy. For this work, we adopt the orbital parallax distance derived from our model as the preferred value.

\subsection{Masses and Stellar Characteristics of $\tau^1$ Aqr} \label{sec:masscharacter}

The masses of the components and the total mass of $\tau^1$ Aqr are calculated using Kepler's laws and the parameters from the radial velocity and interferometric fitting, listed in Table \ref{tab:orbparams}. The stellar masses are $M_1 = 2.56^{+0.057}_{-0.052}$ $M_{\odot}$ and $M_2 = 1.28^{+0.069}_{-0.064}$ $M_{\odot}$, and the total mass of the system is $M_{tot} = 3.84^{+0.089}_{-0.083}$ $M_{\odot}$.

\begin{figure}[ht] 
\centering
    \includegraphics[width=0.8\linewidth]{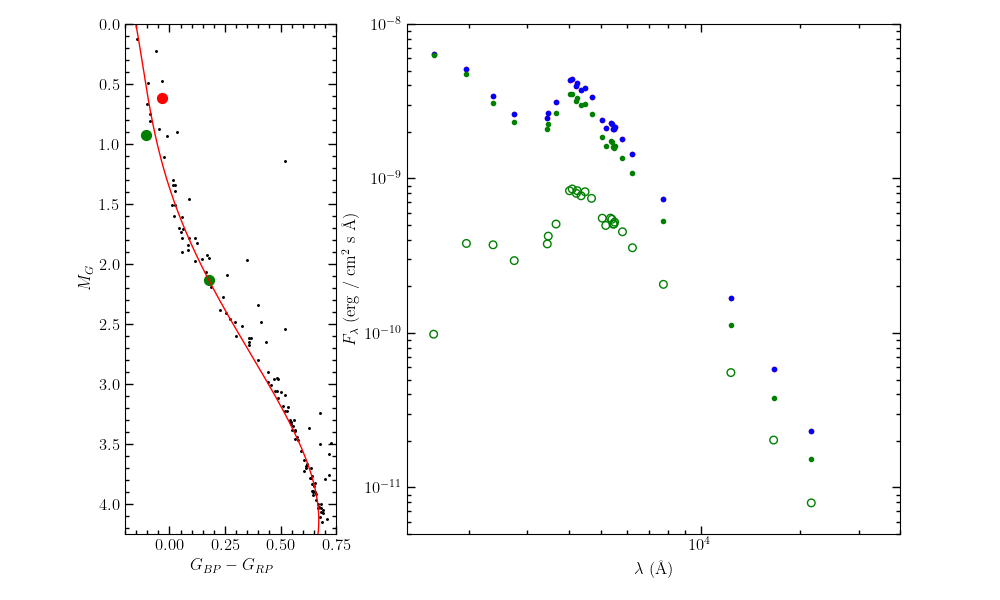}
\caption{CMD and spectral energy distributions (SED) of $\tau^1$ Aqr. \textit{Left panel:} The red line represents the fit of the upper main sequence of Psc-Eri, the \textit{red point} shows the photometry of the $\tau^1$ Aqr binary, and the \textit{green points} show the estimated photometry of the secondary star (based on the interferometric infrared light ratio) and primary star (subtracting the secondary star from the binary light). \textit{Right panel:} photometric SED values for $\tau^1$ Aqr (\textit{blue points}), the primary star (\textit{solid green}) and secondary star (\textit{green circles}).
} 
\label{fig:SED}
\end{figure}

On their own, the masses cannot provide precise age information. To constrain the age, we compare the individual components' positions on the color-magnitude diagram (CMD) against isochrone predictions at their specific measured masses. 
To extract characteristics of the individual components, we use the precise infrared luminosity ratios in the $H$ and $K$ bands along with the system magnitudes $M_H$ and $M_K$ to infer the photometry of the primary and secondary stars, assuming that both components are consistent with the stream's main sequence.
To do this, we collected photometry for luminous members of the Psc-Eri stream $(0 < M_G < 4.25)$ in a wide range of filters from the ultraviolet far into the infrared, corrected for the differing distances using Gaia parallaxes, and fitted the photometry as a function of $M_G$ to map how the flux in each filter changes with position on the main sequence. The photometric catalogs used are shown in Table \ref{tab:sedsources}. The fit for Gaia filters is shown as the red curve in the left panel of Figure \ref{fig:SED}.

\begin{deluxetable}{llcll}
\tabletypesize{\scriptsize}
\tablewidth{0pc}
\tablecaption{Photometry Sources and Flux Calibration for SEDs}
\tablehead{
\colhead{Catalog} & \colhead{Filters} & \colhead{Wavelength} & \colhead{Photometry Ref.} & \colhead{Calibration Ref.}\\
& & \colhead{Range (\AA)} & & }
\startdata
TD1 & F1565,F1965, & $1565-2740$ & \citet{td1uv} & \\
& F2365,F2740 & \\
UVBSSC & spectrophot & $2740-1350$ & \citet{Jamar}& \\
Burnashev & spectrophot. & $3200-7175$ & \citet{burnashev} & \\
 & spectrum & $3300-9400$ & \citet{Rubin}& \\
 Gaia BP/RP   & spectrophot. & $3360-10200$ & & \\
Geneva & $UB_1BB_2V_1VG$ & $3471-5814$ & \citet{geneva} & \\
Str\"{o}mgren & $uvby$ & $3520-5480$ & \citet{paunzen} & \citet{gray}\\
Tycho & $BT,VT$ & $4220-5350$ & \citet{tycho2} & \citet{mvb}\\
Hipparcos & $Hp$ & 5176 & \citet{hippnew} & \citet{mvb}\\
Gaia EDR3 & $G_{BP},G,G_{RP}$ & $5051-7726$ & \citet{gaianear} & \\
2MASS & $JHK_{s}$ & $12350-21590$ & \citet{2mass} & \citet{cohen}\\
WISE & $W1,W2,W3,W4$ & $33526-220883$ & \citet{wise} & \citet{wise}\\
\enddata
\label{tab:sedsources}
\tablerefs{
UVBSSC: UV Bright Star Spectrophotometric Catalog.
2MASS: Two-Micron All-Sky Survey. WISE: Wide-Field Infrared Survey Explorer.}
\end{deluxetable}

The SED in the right panel of Figure \ref{fig:SED} depicts the amount of energy emitted by the binary and both components from the near ultraviolet into the infrared. There is little-to-no difference between the primary and the binary points at the short wavelength end of the spectrum, indicating the dominance of the primary starlight in the ultraviolet and optical. At longer wavelengths, the secondary star contributes more significantly. The fit indicates that the primary is B9 type ($M_G \approx 0.93$, $T_{\rm eff} \approx 11000$K) and the secondary is probably F1 type ($M_G \approx 2.66$, $T_{\rm eff} \approx 7000$ K).
We make use of the resulting photometry for the primary star in an additional SED fitting, discussed in Section \ref{brightstars}, for characteristics of the primary specifically. 

\section{Discussion} \label{sec:discuss}

\subsection{Age Dating} \label{sec:agedate}

\begin{figure}[p] 
    \begin{minipage}{\textwidth}
        \centering
        \includegraphics[scale=0.415]{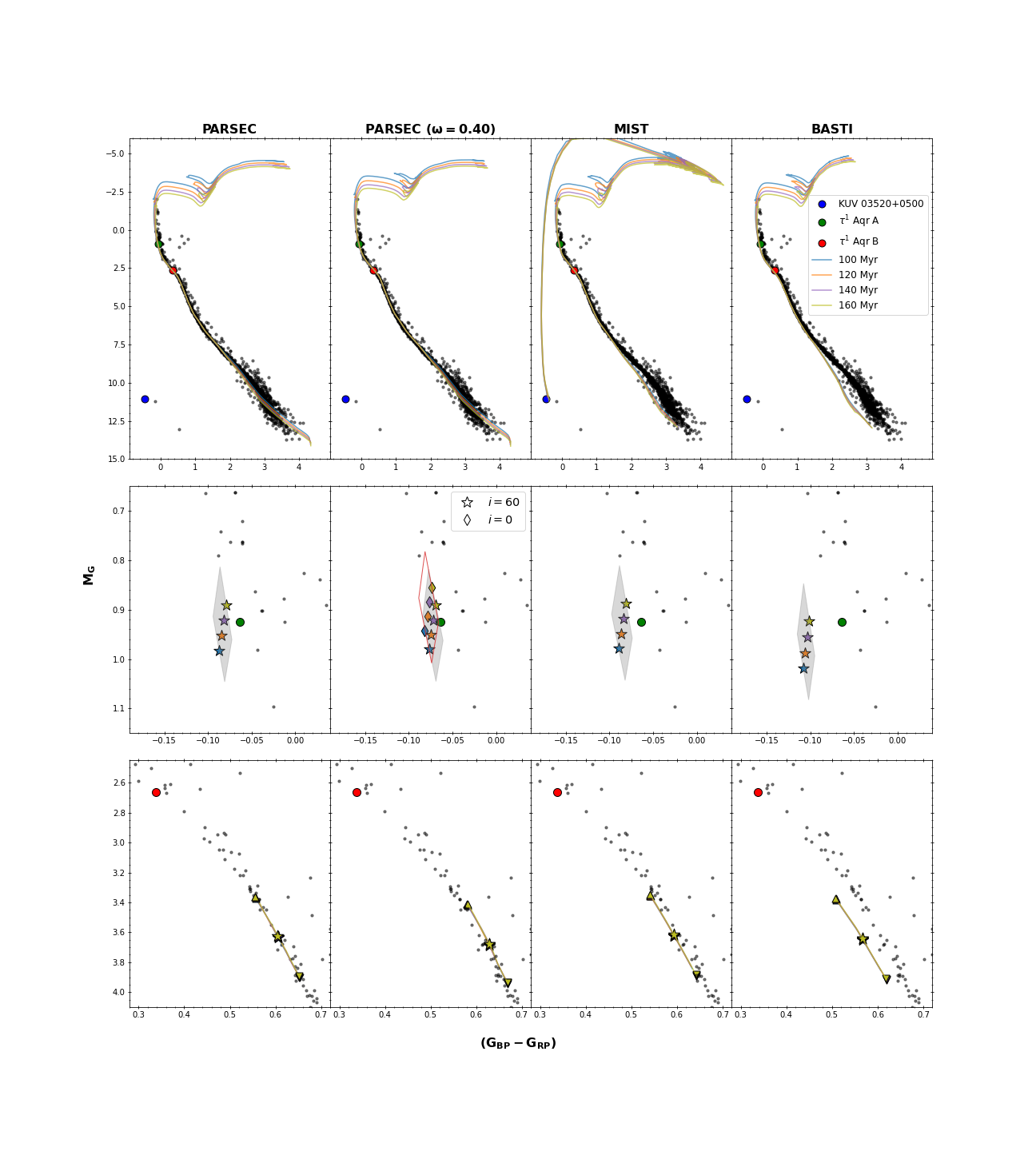}\caption{Comparison of the color-magnitude positions of the binary components to their predicted positions from various isochrone sets for their calculated masses. Each column shows a different isochrone set for the same four ages. \textit{Top row:} Zoomed-out view, including the components of $\tau^1$ Aqr and the candidate WD member. \textit{Middle row:} Zoom on the primary component (\textit{circle}) and the model predicted positions for a star of the same mass (\textit{star markers}). The shaded region illustrates $1\sigma$ mass uncertainties. Additionally, column 2 includes the model predicted positions for the pole-on case (\textit{diamonds}) with a \textit{red outline} for the corresponding uncertainty region. \textit{Bottom row:} Zoom on the secondary component, but with the $1\sigma$ upper and lower mass limits shown with \textit{triangles.}} \label{fig:CMD_ages}
    \end{minipage}
\end{figure}

The calculated masses and photometric measurements of the binary star can be used to constrain the age of the system by comparing to isochrones. The isochrones we utilized are from the PARSEC \citep{PARSECiso},
MIST \citep{MISTiso1,MISTiso2}, and BASTI-IAC \citep{BaSTI_iso} sets.
We interpolate within the isochrones for the predicted absolute magnitude ($M_G$) and color ($G_{BP}-G_{RP}$) for the measured masses of the binary components. The predictions are plotted on the CMDs in Figure \ref{fig:CMD_ages} for comparison with the inferred positions. Although there is agreement with the primary star, all isochrones predict color-magnitude positions about 1 mag fainter than the magnitudes we derived for the secondary star in Section \ref{sec:masscharacter}.

One potential influence on the age estimation is stellar rotation. 
A high rotation velocity is expected to cause bulging at the equator and flattening at the poles, producing an oblate spheroid. This deformation leads to non-uniform surface temperature and gravity, resulting in gravity darkening with higher temperature at the poles and lower temperature at the equator. Aside from these inclination-dependent effects on the observed color and magnitude of the star, there are effects on the evolution timescale. 
The pressure on the core is relieved somewhat because the outer layers are not pressing down as much, and this allows the core to expand and cool slightly, slowing down the rate of nuclear fusion. The star will burn its fuel more slowly, extending its lifetime on the main sequence \citep{fremat_rot}. 

PARSEC isochrones are tabulated with different amounts of initial rotation, parametrized with the non-dimensional rotation rate $\omega$: 
\begin{equation}\label{eqtn:omega}
    \omega = \frac{\Omega}{\Omega_{\rm crit}}, \hspace{5pt} \Omega_{\rm crit} = \sqrt{\frac{GM}{R_{\rm eq}^3}}
\end{equation}
where $\Omega$ is the angular velocity and $\Omega_{\rm crit}$ is the critical angular velocity at which the centrifugal force at the equator is equal to the effective gravity. 
We estimate $\omega = 0.4$ for the primary star, $\tau^1$ Aqr A, using the rotational velocity measured from the broadening function, the inclination and mass determined from the orbital modeling (assuming the stellar spins are aligned with the orbit), and the radius from the SED fitting discussed in Section \ref{brightstars}. Similarly, we estimate $\omega = 0.1$ for the secondary star. The PARSEC rotating stellar isochrones also provide photometric predictions cataloged by different inclination angles. The isochrones that include the primary star's $\omega$ value and the photometric predictions for an inclination angle of 60 deg (closest to the binary system's) are shown in column 2 of Figure \ref{fig:CMD_ages}. The rotation axes for the stars may have aligned with the orbit during formation, but it is unlikely that it would have occurred via tidal interactions after the stars reached the main sequence due to their separation. As a result, it is worth considering the possibility that the spins are not aligned with the orbit, and we include the photometric predictions for a pole-on case, with an inclination angle of 0 deg, in Figure \ref{fig:CMD_ages}. The magnitudes for the pole-on case are brighter by $\sim 0.037$ mag for the $\omega = 0.4$ models, 
and we conclude that rotation and inclination effects have minimal effects on the interpretation of the age of $\tau^1$ Aqr A.

While interstellar extinction cannot explain the magnitude gap of the secondary (as $\tau^1$ Aqr B is brighter than predicted and there are no signs of significant reddening effects in the SED fit of $\tau^1$ Aqr A), there are two other potential sources of error that could affect the agreement. 
First, there could be a significant systematic error in our mass estimates 
and the mass uncertainties can be reduced with additional observations near the extrema of the RV curve. 
The gray areas plotted in the middle row of Figure \ref{fig:CMD_ages} give an indication of the uncertainty in CMD position due to the statistical mass uncertainty. The bottom row also depicts the uncertainty in CMD position of the secondary with triangles representing $1\sigma$ upper and lower mass uncertainties. Although the binary's mass ratio is constrained well by the observations, we have limited radial velocity coverage near periastron and there is a possibility our velocity semi-amplitudes ($K_1, K_2$) are underestimated. If the masses are larger, this could reduce the discrepancy with the secondary star, but could push the primary star out of agreement with the isochrones. For this reason, mass measurements alone are probably not the whole solution to the discrepancy.
Second, uncertainties in the distance to the system could also contribute --- our orbital parallax distance is significantly lower than the Gaia parallax measurement. Because systematic errors in distance would shift both stars equally in absolute magnitude, it is unlikely to eliminate disagreements with models for both stars.

\subsubsection{SED Constraints from Bright Members of Psc-Eri} \label{brightstars}

Additional constraints on the age of the moving group can come from analysis of member star SEDs, which can produce good measures of radii and temperatures. We searched the literature for all known B and A0 star members of Psc-Eri, which are likely to be the most evolved and age-sensitive.
The twelve bright members of Psc-Eri that we examined, along with the primary star of $\tau^1$ Aqr, are listed in Table \ref{tab:brightstars}. 
The stars identified as members by \cite{roser} 
all had a $<$ 1\% probability of belonging to the background population. 
The member stars identified by \cite{ratzen} 
had high ``stability" values of a star in their various classifiers. All tabulated stars had stability $>$ 24\%, except for o Aqr (2.5\%). 
Finally, 
we considered ``high-likelihood candidate members" from MOCA \citep{2026MOCA}, and two lower-probability ``candidate members" ($\kappa$ Lep and 29 Psc). 

For most of the stars (excluding the Be stars), we perform SED fits for the effective temperature $T_{\rm eff}$, reddening $E(B-V)$, and angular diameter $\theta$ by comparing available flux-calibrated spectra and photometry with ATLAS9 models \citep{atlas9model} assuming a fixed surface gravity $\log{g} = 4.0$ (expect for $\tau^1$ Aqr A, which we calculate to have $\log{g} = 4.27$). The parameters $T_{\rm eff}$ and $E(B-V)$ primarily influence the wavelength-dependent shape of a star's spectrum, while $\theta$ affects the total flux level, according to the relation $\theta^2 / 4 = f_{\rm bol} / \sigma T_{\rm eff}^4$. Our fits involved Monte Carlo exploration of the parameter space using the {\tt emcee} package \citep{emcee}. To examine systematic uncertainty introduced by our surface gravity assumption, we also executed the fittings assuming a $\log{g}$ shifted by $\pm 0.25$ and added the resulting offsets in quadrature with the statistical uncertainties for each parameter. The resulting angular diameters are used with published distances\footnote{Distance sources: $\kappa$ Lep, Hipparcos parallax \citep{Hipparcos}; $\tau^1$ Aqr A, orbital parallax from this paper; all others, Gaia parallaxes \cite{BailerJones}.} to determine the radius $(R)$ of each star. The fitted parameters, distances, and radius calculations are listed in Table \ref{tab:brightstars}, along with their projected rotational velocities\footnote{The values listed are the medians of the values cataloged in \cite{Glebocki}. For $\kappa$ Lep, a value from \cite{Zorec} was also included in the median. Values for o Aqr and $\beta$ Psc are from \cite{Arcos}. For $\tau^1$ Aqr A, the value was derived in Section \ref{sec:RVspec}.} $(v_{\rm rot} \sin{i})$, rotation rate $(\omega)$ estimates, the spectrophotometry data used, and their membership sources. More information on these stars can be found in Appendix \ref{app:sed}.

\begin{deluxetable}{llccccccccc}
\tablewidth{0pc}
\tablecaption{Bright Stars of the Psc-Eri Stream \label{tab:brightstars}}
\tabletypesize{\footnotesize}
    \tablehead{ \colhead{Name} & \colhead{Sp Type} & \colhead{$d$ (pc)} & \colhead{$T_{\rm eff}$ (K)} & \colhead{$E(B-V)$} & \colhead{$\theta$ (mas)} & \colhead{$R$ $(R_\odot)$} & \colhead{$v\sin{i}$ (km/s)} & \colhead{$\omega$} & \colhead{Spectrophotometry } & \colhead{Membership}} 
\startdata
$\kappa$ Lep & B7III+B9V   & $223 \pm 28.9$ & $12426 \pm 57$ & $0.0224^{+0.0028}_{-0.0029}$ & $0.338 \pm 0.001$ & $8.11 \pm 0.05$ & 122.5 & 0.36 & 3, 4, 7 & C, D\\
o Aqr\tablenotemark{a}                         & B5Ve        & $143^{+2.8}_{-3.1}$ & $13080\pm30$        & $0.022\pm0.003$ & $0.245\pm0.001$ & $3.60\pm0.02$ & 350 & 0.95 & 2, 3, 4, 9& A, C, D\\
$\beta$ Psc\tablenotemark{a}                   & B6Ve        & $124^{+1.9}_{-1.8}$ & $14350\pm60$ & $0.024\pm0.001$ & $0.266\pm0.002$ & $3.64\pm0.03$ & 90 & 0.64 & 2, 3, 5, 9  & C\\
29 Psc                        & B7III-IVpHe-wkMn & $131^{+2.9}_{-3.0}$ & $13655 \pm 100$ & $0.0000$ & $0.219 \pm 0.001$ & $3.08 \pm 0.03$ & 67 & 0.13 &  1, 5, 6, 8 & C, D\\
$\nu$ For                     & B9.5IIIspSi & $106^{+1.4}_{-1.2}$ & $12413 \pm 41$ & $0.0000$ & $0.291 \pm 0.001$ & $3.32 \pm 0.06$ & 45 & 0.16 &  1, 3, 7 & A, D\\
2 Cet\tablenotemark{a}                         & B9IVne       & $95 \pm 0.8$  & $10520\pm20$ & $0.020\pm0.001$ & $0.346\pm0.001$ & & 200 & 0.77 & 1, 2, 3, 5, 9 & C\\
106 Aqr\tablenotemark{a}                       & B9Vn        & $132^{+5.6}_{-4.9}$ & $11480 \pm 25$ & $0.000^{+0.002}$ & $0.240 \pm 0.005$ & $3.45 \pm 0.10$ & 310 & 0.9 &  1 & A, D\\
108 Aqr                       & A0VpSiSrCr  & $100 \pm 1.5$ & $11598 \pm 41$ & $0.0001 \pm 0.0001$ & $0.249 \pm 0.001$ & $2.68 \pm 0.07$ & 30 & 0.07 & 3  & A, B, C, D\\
HD 1064                       & B8.5V       & $108 \pm 0.8$ & $11687^{+40}_{-39}$ & $0.0256^{+0.0025}_{-0.0026}$ & $0.199 \pm 0.001$ & $2.32 \pm 0.06$ & 124 & 0.25 & 1  & A, C, D\\
$\tau^1$ Aqr A & B9V         & $103 \pm 1.1$ & $11383 \pm 46$ & $0.0001 \pm 0.0001$ & $0.175 \pm 0.001$  & $1.95 \pm 0.06$ & 143 & 0.35 & 1,10  & A, B, C, D\\ 
26 Psc       & B9V         & $121^{+0.9}_{-0.7}$ & $11532^{+76}_{-77}$ & $0.0376 \pm 0.0028$ & $0.162 \pm 0.001$ & $2.11 \pm 0.02$ & 25 & 0.05 & 7  & A, C, D\\
HD 218639    & A0V         & $103 \pm 0.3$ & $9862 \pm 99$  & $0.0580 \pm 0.0088$ & $0.185 \pm 0.001$ & $2.05 \pm 0.02$ & 235 & 0.49 & -  & A, B, C, D\\
HD 209105    & A0          & $180^{+1.5}_{-1.4}$ & $8794\pm 51$  & $0.0003^{+0.0007}_{-0.0003}$ & $0.091 \pm 0.001$ & $1.77 \pm 0.05$ & - & 0.0 & 7 & A, B, C, D\\
\enddata 

\tablenotetext{a}{FASTROT model fits, with volumetric temperatures and radii and fitted $\omega$. }
\tablecomments{\textbf{Spectrophotometry References.—} 1. TD1, 2. IUE, 3. \citet{burnashev}, 4. UVBSSC \citep{Jamar}, 5. UVBSSS \citep{uvbsss}, 6. \citet{Rubin}, 7. Gaia BP/RP, 8. HST \citep{Pal}, 9. \citet{cochetti}, 10. Our photometry of the primary component that was obtained from the disentangled SED described in Section \ref{sec:masscharacter}. \textbf{Membership References.—} A. \citet{MAINcurt}, B. \citet{roser}, C. \citet{ratzen}, D. MOCA \citep{2026MOCA}.}
\end{deluxetable}

We make an effort to incorporate the two Be stars, o Aqr and $\beta$ Psc, into our analysis because they are some of the most age sensitive members in the group. These Be stars require a different approach, as they are rapidly rotating with emission lines previously detected in their spectra. The emission is thought to arise from a circumstellar disk of ionized gas containing material lost from the equator, as discussed in Section \ref{sec:agedate} \citep{rivrev}. 
This, along with gravity darkening, can skew the $T_{\rm eff}$ inferred from spectra. Depending on the inclination of the star, the observed spectrum may reflect a temperature closer to that of the cooler equatorial region or the hotter polar region. Similarly, rapid rotation affects apparent surface gravity ($\log g$).

For these Be stars, we conducted fits using FASTROT models \citep{montesinos}. These use the Roche model to describe the surface gravity and temperature variations as a function of latitude on the rotationally-distorted photosphere, where the emission from each small patch is calculated from an ATLAS9 model with the appropriate characteristics. In our fits, the temperature parameter is the polar temperature $T_{\rm pole}$ and the angular diameter information is split into an equatorial radius parameter $R_{\rm eq}$ and a distance (using a Gaussian prior). In addition to those parameters, we fit for the inclination of the spin axis using $\cos i$ as a parameter, and for the angular rotation rate $\omega = \Omega / \Omega_K$ that determines the amount of shape distortion in combination with an assumed stellar mass. We used measured projected rotation speeds $v_{\rm rot} \sin i$ as constraints in the likelihood calculation. 

For $\beta$ Psc, we use a high-resolution IUE spectrum from the INES database (SWP15935HL; \citealt{INES}) in the ultraviolet to help constrain the reddening. This spectrum was taken with the large science aperture for reliability of the absolute flux calibration \citep{INESabs}. We also used a flux-calibrated CASLEO spectrum using the Boller \& Chivens Cassegrain spectrograph attached to the Jorge Sahade telescope, as presented in \citet{cochetti}. The PM 512 detector gave an effective resolution of 4.53 \AA \: every two pixels $(R \approx 900)$ and a spectral range $3500–4600$ \AA. For o Aqr, we merged two SWP and two LWP low-resolution IUE spectra after correcting the fluxes according to the procedure of \citet{mf2000}.  For 2 Cet, we also used a merged low-resolution IUE spectrum computed from two LWR and six SWP observations. For both o Aqr and 2 Cet, we utilized flux-calibrated CASLEO spectra taken with the 600 l/mm grating and a slit width of 250 $\mu$m. The covered spectral range was $3500–5000$ \AA \: with effective resolution 5.25 \AA \: every two pixels $(R \approx 660)$.

For these Be-type stars, we quote volumetric radii and temperatures in Table \ref{tab:brightstars}. The volumetric radius $R_{\rm vol}$ is the radius of a spherical star that would give the same volume as the rotationally-distorted star, and the volumetric temperature $T_{\rm vol}$ is calculated from $R_{\rm vol}$ and the inferred total luminosity. The quoted angular diameter $\theta$ is also a volumetric average that will differ from measurements, but is representative for the star. We note that the best-fit angular rotation rate for $\beta$ Psc is well below critical rotation and the inferred spin axis inclination is low at $i \approx 19\pm1^\circ$. o Aqr is inferred from our fits to be rotating much closer to break-up ($\omega = 0.87$) with a much higher inclination  of $63\pm2^\circ$.
These stars are discussed in additional detail in Appendix \ref{app:sed}.

\begin{figure}[ht] 
\centering
    \includegraphics[width=1\linewidth]{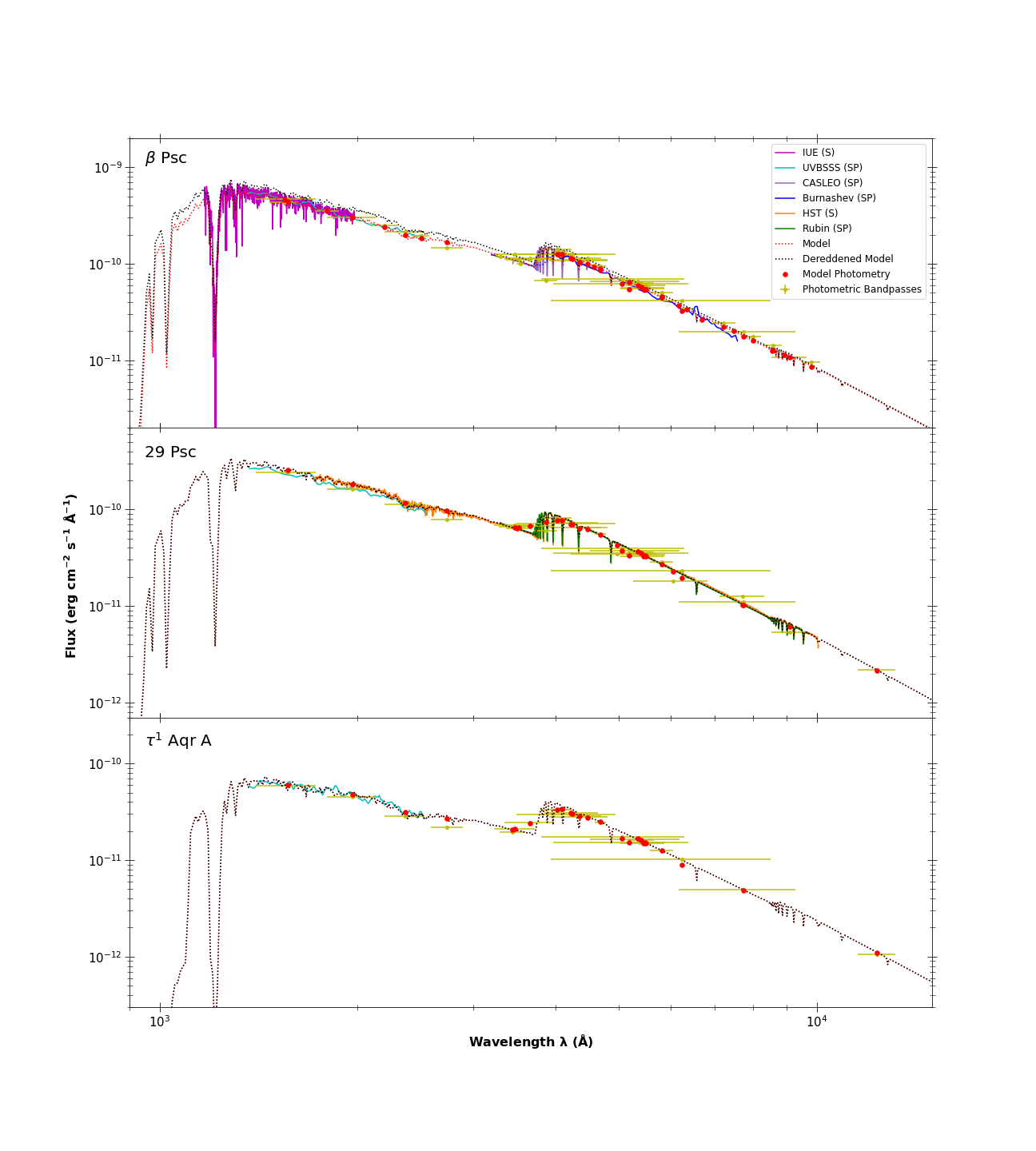}
\caption{SED fitting examples. The \textit{red dotted line} is the best-fit model spectrum with reddening, and the \textit{black dotted line} is the unreddened model spectrum. \textit{Yellow points} are photometric measurements, with \textit{horizontal bars} representing the bandpasses of the filters. Model photometry is shown with \textit{red points}. Spectroscopic (S) and spectrophotometric (SP) data are shown with \textit{solid lines}, as specified in the legend. A portion of the spectrophotometry ($3300-4000$ \AA, and $> 10000$ \AA) for $\beta$ Psc from Burnashev is shown with a \textit{dotted line} because it was not used in the fitting, as explained further in Appendix \ref{app:sed}.}
\label{fig:all3SEDs}
\end{figure}

To illustrate more clearly how effective temperature and reddening influence spectral shapes, we present examples of SED fittings for $\beta$ Psc, 29 Psc, and $\tau^1$ Aqr A in Figure \ref{fig:all3SEDs}. The shape of the spectrum varies considerably with the temperature of these three stars, especially in the ultraviolet (UV) range. The coolest of the three, $\tau^1$ Aqr A, has the steepest drop-off in the UV. We can also observe the effects of modest reddening in the spectrum of $\beta$ Psc.

\begin{figure}[p] 
    \begin{minipage}{\textwidth}
        \centering
            \includegraphics[scale=0.44]{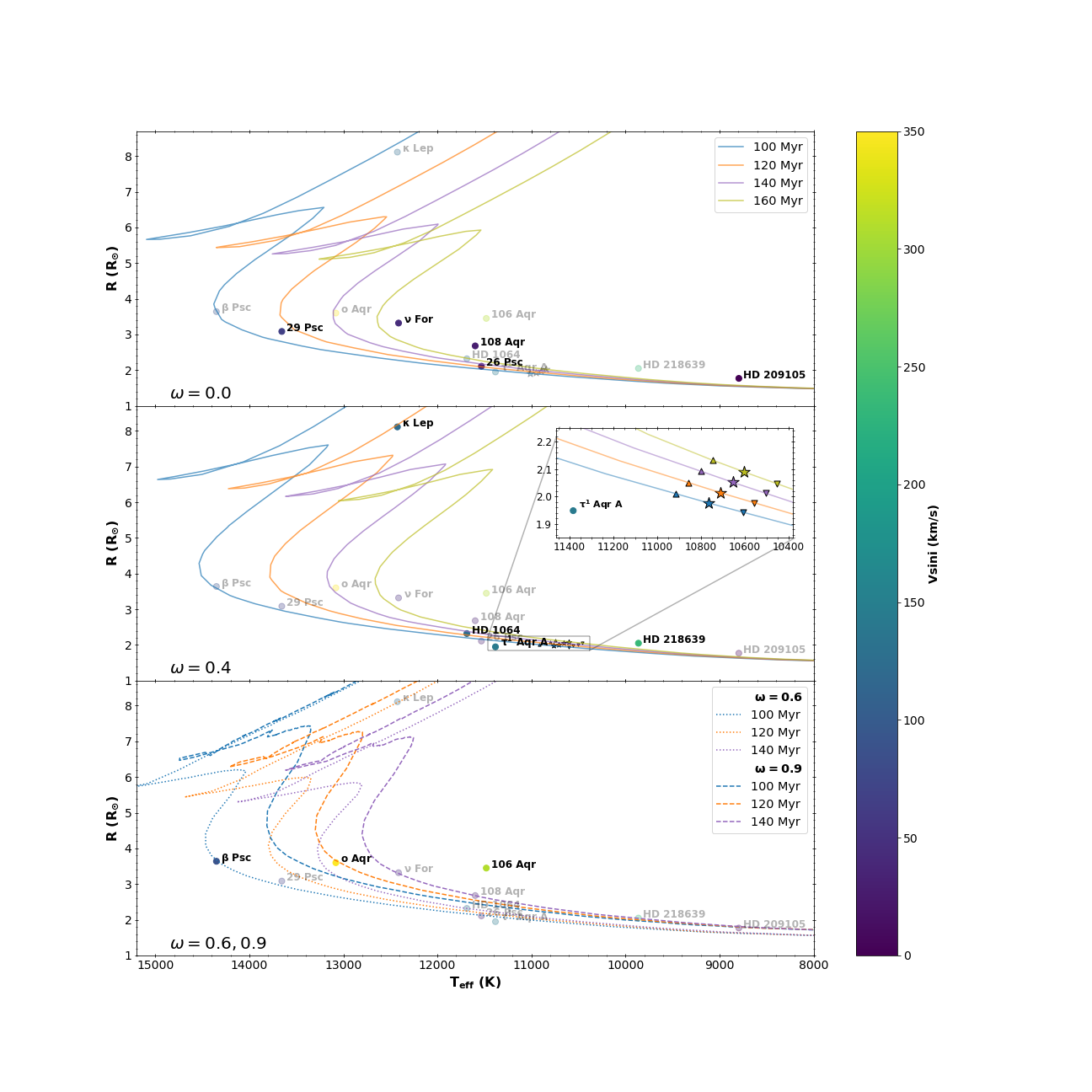} \caption{Radius versus effective temperatures for bright star members. Stars are plotted with emphasis in the rotation rate bin indicated by measurements. \textit{Top panel:} Slow-rotating stars, compared with non-rotating ($\omega = 0.0$) PARSEC isochrones. \textit{Middle panel:} Medium-rotating stars, compared with PARSEC isochrones with $\omega = 0.4$. Points on the isochrones denoted by \textit{star markers} represent predictions for the $R$ and $T_{\rm eff}$ of a star of $\tau^1$ Aqr A's mass. The \textit{upright/inverted triangles} are the predicted positions for the maximum/minimum masses determined from the mass uncertainties. \textit{Bottom panel:} Fast-rotating stars, compared with SYCLIST isochrones with $\omega = 0.6, 0.9$. 
            }
            \label{fig:RvsTeff}
    \end{minipage}
\end{figure}

\begin{figure}[ht!] 
\centering
    \includegraphics[width=0.6\linewidth]{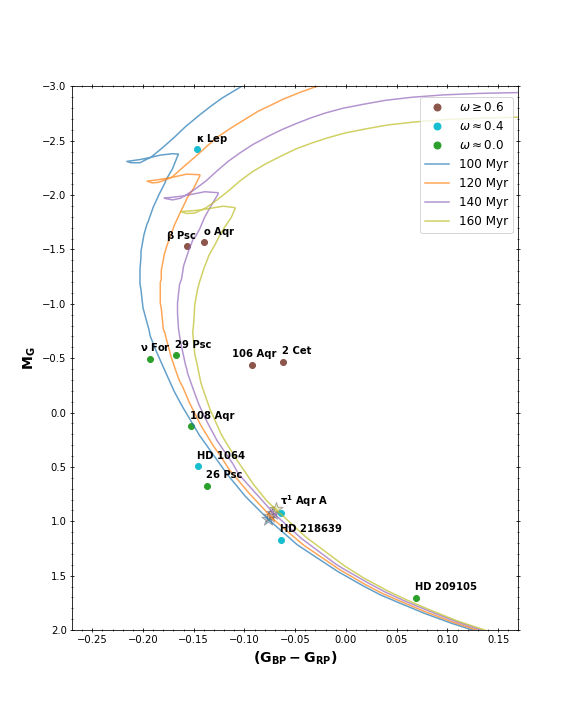}
\caption{Color-Magnitude Diagram (CMD) of the 12 bright members. The stars are organized into 3 groups, $\omega \approx$ 0, 0.4, and $\omega \geq 0.6$ , based on their estimated rotation rates. The $E(B-V)$ values from the SED fittings were used to deredden the CMD. The Be stars, $\beta$ Psc and o Aqr, are corrected for their viewing angle using the FASTROT models. The isochrones are from PARSEC with a rotation rate of $\omega = 0.4$.}
\label{fig:brightsCMD}
\end{figure}

We plot the effective temperatures and radii of the bright stars in Figure \ref{fig:RvsTeff}, along with PARSEC and SYCLIST \citep{SYCLIST} isochrones with varying rotations. For the slow to intermediate rotating stars, we use PARSEC isochrones of ages $100-160$ Myr with rotation rates of $\omega =$ 0 and 0.4. PARSEC uses an ellipsoid approximation to equipotential surfaces for rotating stars \citep{EndalSofia} that is valid for $\omega \lesssim 0.6$, while a Roche potential better models stars with $\omega$ up to approximately 0.9 \citep{Torres}. For that reason, we use SYCLIST isochrones for stars with $0.6 < \omega < 0.9$. We have estimated $\omega$ for each star using Equation \ref{eqtn:omega}. The radii from our fittings and estimated mass values are used to determine $v_{\rm crit}$, and $v$ is derived from the $v\sin{i}$ of each star assuming an equator-on spin inclination $(i = 90^\circ)$, for all stars except those with known inclinations ($\nu$ For, 108 Aqr, and $\tau^1$ Aqr A) or Be stars with constraints from SED fitting.

We find $\tau^1$ Aqr A to be about 5-7\% hotter and with a radius about 1-6\% smaller than the PARSEC $\omega = 0.4$ isochrone model predictions. The isochrones predict a luminosity within the range $47.8 - 50.5$ $L_\odot$, while our SED calculations return a higher luminosity of $57.4$ $L_\odot$ for $\tau^1$ Aqr A. If we use MIST isochrones with $\omega = 0.4$, we find similar discrepancies in radius, temperature, and luminosity. We considered the effects of different metallicities in an effort to resolve the discrepancy, and find that a slightly sub-solar metallicity of [Fe/H]$=-0.09$ would bring the primary into better agreement in Figure \ref{fig:RvsTeff}. This metallicity is plausible given that it falls within the measurement uncertainties and dispersion of Psc-Eri reported in \cite{metalpsceri}. However, adopting a lower metallicity would shift the CMD comparison and would not improve the offset with the secondary component, so we proceed with the solar-metallicity tracks that are plotted.

We plot the 12 bright stars in the CMD in Figure \ref{fig:brightsCMD}, showing the same $\omega$ binning as used in Figure \ref{fig:RvsTeff}. The CMD illustrates the stars nearest to the main sequence turnoff point of Psc-Eri, and therefore the most age sensitive, are $\kappa$ Lep, o Aqr, $\beta$ Psc, 29 Psc, and $\nu$ For. However, we provisionally remove $\nu$ For from our analysis because of the possibility that it may be a binary star, as discussed in Appendix \ref{app:sed}. The positions of $\kappa$ Lep, o Aqr, $\beta$ Psc, and 29 Psc in Figure \ref{fig:RvsTeff} align with ages ranging from 100 to 120 Myr.

\subsection{The White Dwarf Initial-Final Mass Relation} \label{sec:WD}

Although the age for Psc-Eri is still in some doubt, we utilized our age range to calculate the range of possible initial mass values for the WD member KUV 0352+0500. Temperature measurements for the white dwarf give values near 37000 K. [Spectroscopic measurements give $36920\pm620$ K \citep{gianninas} and $37250\pm125$ K \citep{koester} from Balmer line fits, while \citet{torres_wd} get 34580 K from fitting to photometry based on Gaia XP spectra. Because available spectrophotometric measurements are all on the Rayleigh-Jeans tail of the spectral energy distribution for this hot WD, we do not regard the photometric temperature to be as trustworthy.] Older models \citep{limoges} give a WD cooling age of 60 Myr, while a $1.1 M_\odot$ ultramassive ONe WD cooling model from \citet{camisassa} gives a cooling age of 35 Myr, and a $1.1 M_\odot$ CO WD cooling model from \citet{bedard} gives a cooling age of 40 Myr. A relatively large uncertainty of 500 K in the temperature leads to an uncertainty of around 3 Myr in the cooling age, comparable in size to the differences between cooling models.

\begin{figure}[ht!] 
\centering
    \includegraphics[width=0.9\linewidth]{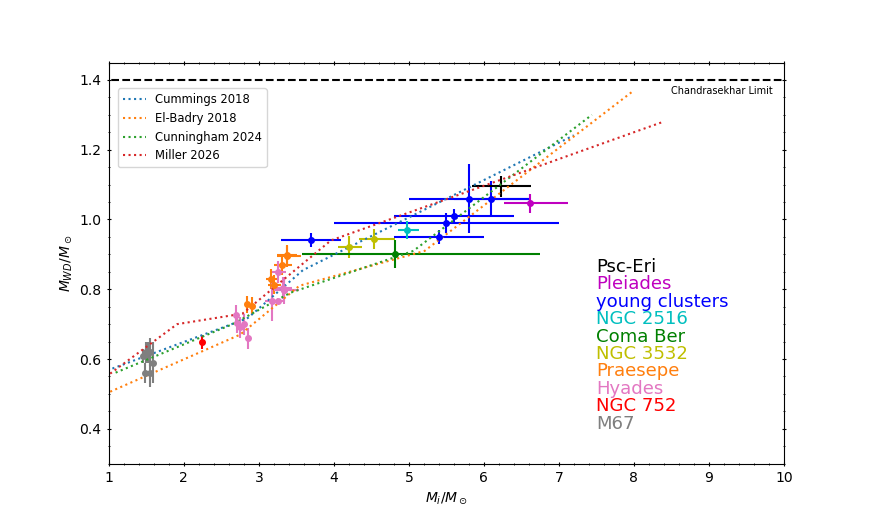}
\caption{ The white dwarf IFMR for various clusters. Psc-Eri's WD member KUV 0352+0500 is plotted in black using values found in this work. For the other clusters, the data sources are: Pleiades, \cite{Heyl}; young clusters, \cite{Richer}; NGC 2516, NGC 3532, and Hyades, \cite{Cummings2018}; Coma Ber, \cite{Dobbie} (WD) and \cite{Lam} (initial mass); Praesepe, \cite{Cummings2018} (WD) and \cite{Morales} (initial mass); NGC 752, \cite{Marigo}; and M67, \cite{Canton}. The \textit{dotted lines} represent piecewise IFMR fits to field stars \citep{ElBadry, Cunningham} and clusters \citep{Cummings2018,miller26}. }
\label{fig:WDIFMR}
\end{figure}

We therefore use nonrotating PARSEC isochrones and a WD cooling age of $40\pm7$ Myr.
Assuming an age of 100 Myr yields an initial mass of $M_i = 6.63$ $M_\odot$, while an age of 120 Myr indicates $M_i = 5.84 M_\odot$. We used these results and the average of the WD mass measurements from \cite{gianninas} and \cite{limoges}, $M_{WD} = 1.095\pm 0.03$ $M_\odot$, to place KUV 0352+0500 on the IFMR plot in Figure \ref{fig:WDIFMR}. Here, the initial mass error bar represents the initial mass range of $5.84-6.63$ $M_\odot$ implicated by our age range. While there is more work to be done to improve the precision of the measurements, KUV 0352+0500 agrees well with the four IFMR fits shown in Figure \ref{fig:WDIFMR}.

\section{Conclusions} \label{sec:conclusion}

We successfully modeled the orbit of the binary system $\tau^1$ Aqr, obtaining precise orbital parameters derived from spectroscopic and interferometric data. These orbital parameters were used to determine masses of both component stars: $M_1 = 2.56^{+0.057}_{-0.052}$ $M_\odot$ and $M_2 = 1.28^{+0.069}_{-0.064}$ $M_\odot$. 
We fitted the individual component SEDs derived from our measured infrared luminosity ratios and the stream's main-sequence siblings. We find that the secondary star is roughly 1 mag brighter than models predict for the measured masses. For more perspective on the age of the stellar stream, we analyzed the radii and temperatures obtained through SED fits of other bright stars in the stream and compared these observations to isochrones as well. This suggests an age in the range of $100-120$ Myr, consistent with the ages determined by \cite{MAINcurt} and \cite{roser}. 

Our orbital parallax distance for $\tau^1$ Aqr ($d = 103 \pm 1.1$ pc) is significantly smaller than the Gaia parallax distance ($109.01$ pc). An improved Gaia astrometric solution for $\tau^1$ Aqr that accounts for the effects of its binarity would help us better understand the seriousness of the disagreement in the secondary star's magnitude. The spatial extent of the Psc-Eri stream makes distance measurements critical to understanding its stellar population. If the stream formed with separate centers of star formation as suggested by \citet{MAINcurt}, it might contain stars that are somewhat older (and thereby larger and more luminous) than expected, although it is unlikely that this can explain the magnitude discrepancy on its own. Model physics uncertainties such as convective core overshooting and rotation might also be contributing to the disagreement, but do not seem capable on their own of explaining the full size of it.

Our preliminary analysis of the massive WD KUV 0352+0500 in the Psc-Eri stream demonstrates how significant a precise age for KUV 0352+0500 could be in improving our comprehension of the WD IFMR. Should a precise age be obtained, the resulting mass estimate of the progenitor star would further improve constraints for the least understood high-mass end of the relation.

\begin{acknowledgments}
We gratefully acknowledge support from the National Science Foundation under Grant No. AAG 1817217 to E.L.S.
This work is based upon observations obtained with the Georgia State University Center for High Angular Resolution Astronomy Array at Mount Wilson Observatory.  The CHARA Array is supported by the National Science Foundation under Grant No. AST-2034336 and AST-2407956. Institutional support has been provided from the GSU College of Arts and Sciences, Office of the Provost, and Office of the Vice President for Research and Economic Development. 
Time at the CHARA Array was granted through the NOIRLab community access program (NOIRLab PropID:2023B-745281, 2024B-297396; PI: E. Sandquist).
SK acknowledges funding for MIRC-X received funding from the European Research Council (ERC) under the European Union's Horizon 2020 research and innovation programme (Starting Grant No. 639889 and Consolidated Grant No. 101003096) and from STFC Small Award (ST/Y002695/1). JDM acknowledges funding for the development of MIRC-X (NASA-XRP NNX16AD43G, NSF-AST 2009489) and MYSTIC (NSF-ATI 1506540, NSF-AST 1909165).
This research has used observations from the SIMBAD database, operated at CDS, Strasbourg, France; and from the Montreal Open Clusters and Associations (MOCA) database, operated at the Montr\'eal Plan\'etarium \citep{2026MOCA}. The research has also made use of the Jean-Marie Mariotti Center OiDB (available at http://oidb.jmmc.fr), Aspro, and SearchCal services.
\end{acknowledgments}

\facilities{CHARA (MIRC-X,MYSTIC,PAVO), VLTI (PIONIER), OCA:Hexapod (BESO), VLT:Kueyen (X-Shooter,UVES)}

\bibliography{TheBib}{}
\bibliographystyle{aasjournalv7}

\clearpage

\appendix
\section{Notes on Bright Psc-Eri Stream Members}\label{app:sed}

{\it $\kappa$ Lep:} This is likely to have the largest star by radius in the stream, but it is also a visual binary. The primary is a B7III star \citep{gg}, while the companion may be a B9 star \citep{gahm,ge2024} about 2.5 mag fainter at $2^{\prime\prime}$ from the primary (with decreasing separation since its discovery in 1832). The companion should have a small effect on the SED and the calculation of a radius, but it complicates the distance measurement. Hipparcos produced a  nine-parameter double-star solution for the system having a parallax of $5.83\pm0.69$ mas. On the other hand, the Gaia DR3 astrometric solution is potentially affected by the companion and Gaia's parallax ($6.98\pm0.19$ mas) may be affected, as it has a somewhat elevated RUWE of 1.450. A precise understanding of the star's characteristics may need to wait for the clarity of a binary star fit to 
Gaia data, but we believe the Hipparcos parallax is currently more reliable. This gives $\kappa$ Lep a significantly larger radius than would be computed using the Gaia parallax. $\kappa$ Lep B is given as a candidate stream member in MOCA \citep{2026MOCA}.

{\it 106 Aqr:} This star falls at around the luminosity of the main sequence turnoff, but we find it to be significantly redder than other stars. A number of pieces of circumstantial evidence indicate it may be a binary. 106 Aqr is unusual for its detection in X-rays \citep{Xrayberg}, which sometimes can indicate a late-type stellar companion. No such companion has been identified to date though \citep{hubrig}. 106 Aqr is identified as a variable from TESS observations, with disagreement on the type (ellipsoidal, rotational, or slowly-pulsating B variable, \citealt{pedersen}; ellipsoidal, \citealt{balona19}). It was found to have constant radial velocity by \citet{BESOspect}, but significant velocity variability during one night by \citet{baade}. These clues, along with the unusually high radius found from our SED fitting, may point toward a close binary, but the evidence is not yet persuasive. The large $v_{\rm rot} \sin i$ for the star (probably in excess of 300 km s$^{-1}$; \citealt{huang,alg}) recommends that it be put in the rapidly rotating category for the Psc-Eri stream, in spite of the fact that it is not known to be a Be star. However, before it is used further to characterize the stream, the possible binarity needs to be tested further.


\subsection{Be Stars}

{\it o Aqr:} o Aqr was discussed most recently by \citet{almeida}, and is a Be star with a stable H$\alpha$ emission line profile with central absorption that implies a fairly high spin inclination. Because of the likely effects of a circumstellar disk, we omitted flux measurements around the Balmer jump ($330-400$ nm) and longer than 800 nm. Our fits to the SED push for an inclination  of $49\pm3^\circ$, disagreeing with the H$\alpha$ and Br$\gamma$ interferometric measurements ($76^\circ$ and $61^\circ$) of \citet{almeida}. (\citealt{cochetti19} found $i=70\pm5^\circ$ with the same Br$\gamma$ data, and \citealt{sigut15} found $75\pm3^\circ$ from H$\alpha$ interferometric and spectroscopic analysis of the circumstellar disk.) \citeauthor{almeida} looked for signs of ellipticity in the star using CHARA VEGA observations in the continuum around the H$\alpha$ line, but found that the observations were consistent with a uniform disk having $\theta = 0.28\pm0.01$ mas. 

\begin{figure}[ht!] 
\centering
    \includegraphics[width=0.9\linewidth]{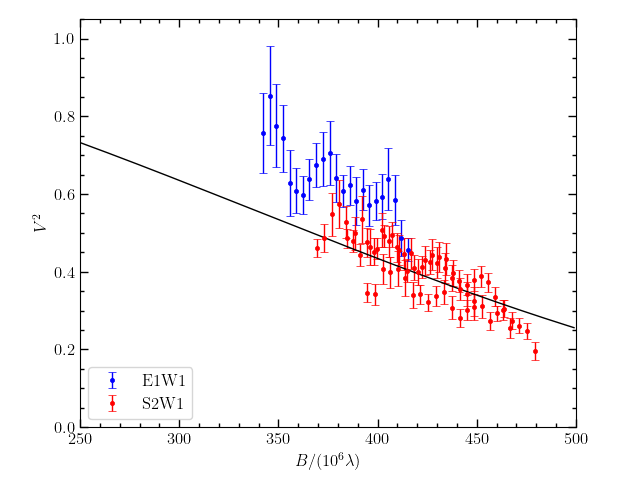}
\caption{ Visibilities measured for o Aqr using the PAVO beam combiner at CHARA. The \textit{solid line} shows a uniform disk model with angular diameter $\theta = 0.295$ mas with a limb-darkening radius correction.
}
\label{oaqrpavo}
\end{figure}

We processed two nights of archival CHARA PAVO observations for o Aqr (PI D. Gies; see Table \ref{tab:ointerfdata}) covering red and near-infrared wavelengths using {\tt pavo\_idl} software on the Georgia State data reduction computer.  We fitted the visibilities with {\tt pavo\_idl}, and found a limb-darkened diameter $\theta = 0.295\pm0.003$ mas (see Fig. \ref{oaqrpavo}) consistent with the earlier observations. Although the two nights of observations appear to imply different angular sizes (as seen in the vertical offset in the visibilities), the observations were taken essentially all at the same orientation (position angle $75^\circ$), and so probably do not reflect ellipticity of the star.
The lack of detected ellipticity in earlier observations is puzzling given the star's rapid rotation and fairly high spin inclination, as stars rotating near break-up could have as much as a 30\% variation as a function of position angle. Our SED-based volumetric radius is in agreement with the \citeauthor{almeida} interferometry measurement ($4.0\pm0.3$ $R_\odot$), although it is not clear whether their quoted value accounts for ellipticity at all. We note that residual effects of circumstellar material on the angular diameter measurements would be likely to inflate the radius, and so age inferred from o Aqr is likely to be somewhat overestimated. Further observations would help address these questions.


\begin{deluxetable}{ccccclc}
\tabletypesize{\scriptsize}
\tablewidth{0pc}
\tablecaption{Interferometric Observations of o Aqr \label{tab:ointerfdata}}
    \tablehead{\colhead{UT Date} & \colhead{mJD\tablenotemark{a}}  &  \colhead{Combiner} & \colhead{$N_{tel}$} &  \colhead{$N_{bracket}$}  & \colhead{Calibrators} & \colhead{$\theta$ (mas)}}
    \startdata
      2014 Aug 6  & 60201.69169 &  CHARA-PAVO   & 2 & 1 & HD 210424, 211924 & $0.246\pm0.007$ \\
      2014 Aug 7  & 60205.72421 &  CHARA-PAVO   & 2 & 3 & HD 210424, 211924 & $0.300\pm0.003$\\ 
    \enddata
\tablenotetext{a}{mJD = HJD $- 2400000$}
\end{deluxetable}

{\it $\beta$ Psc:} This star with strong H$\alpha$ emission was discussed by \citet{levenhagen}. The ``bottle-shaped" H$\alpha$ profile is indicative of a low inclination, face-on disk. The Balmer jump for the system appears to be depressed in size for a star of its type, as can be seen in the spectrophotometry of \citet{burnashev} and \citet{cochetti}, as well as a low Str\"{o}mgren $c_1$ photometric index. This is likely contamination by scattered light from circumstellar material. For this reason, in our SED fitting, we have omitted photometric and spectrophotometric data surrounding the jump in the range of $330-400$ nm, as well as disk-contaminated wavelengths above 1000 nm. Our fits assumed a mass of $4.3M_\odot$, and used a high-resolution spectrum from IUE, UV spectrophotometric data from \citet{uvbsss}, and part of the \citet{cochetti} spectrophotometry. Our results are given in Table \ref{tab:brightstars}, and are in reasonable agreement with the results of \citet{levenhagen}, who also used rotationally-distorted surface models to fit a more limited set of optical and infrared photometry but with detailed modeling of line profiles in high-resolution spectra. Table \ref{tab:bPsc} compares the results for some of the fit parameters.

\begin{deluxetable}{lcc}
\tabletypesize{\scriptsize}
\tablewidth{0pc}
\tablecaption{Best-Fit Models for $\beta$ Psc}\label{tab:bPsc}
\tablehead{
\colhead{Parameter} & \colhead{This Work} & \colhead{Levenhagen et al.} }
\startdata
$T_{\rm pole}$ (K) & $15190\pm100$ & $15556\pm600$\\
$R_{\rm eq} / R_\odot$ & $3.77\pm0.01$ & $4.09\pm0.21$\\
$\omega$ & $0.45\pm0.05$ & $0.79\pm0.13$\\
$i$ (deg) & $25\pm3$ & $20\pm3$\\
\enddata
\end{deluxetable}

{\it 2 Cet:} This is a star near the main sequence turnoff that has been sporadically rediscovered as a Be star \citep{harmanec,cote,Beatlas}. It was reclassified from Bn to Be by \citet{cochetti} after finding a shell-like emission component to the H$\alpha$ line with a shape implying an intermediate inclination, and we adopt a revised spectral type here. We find, in agreement with \citealt{cochetti}, that the SED shows a large Balmer decrement indicating absorption due to a circumstellar disk. This is supported by photometric indices that are sensitive to the Balmer jump, such as Str\"{o}mgren $c_1$ and the $d$ index in the Geneva system, where 2 Cet presents large values compared to other B stars. 
\citet{bando} classified the TESS light curve as showing rotational variability and Algol-type (EA) eclipses. The detrended light curve does show variations  that resemble ellipsoidal variations, but at a low amplitude of 65 ppm and an implied orbital period of about 1.39 d \citep{fetherolf}. The star is brighter and redder than the other group members in the Gaia CAMD, which also supports the idea of a bright companion. However, the radial velocity was found to be constant by \citet{BESOspect}, and no visual companion has been found with adaptive optics \citep{roberts}. This could be explained if both stars in a binary were rapidly rotating, so that their lines usually were not distinguishable. The system deserves additional study, and for this reason, we have not quoted a radius for the star.

\subsection{Chemically Peculiar Stars}

Chemically peculiar stars are frequently found to have low $v_{\rm rot} \sin i$ \citep{aandm}, and genuinely slow rotation allows chemical peculiarities to appear and survive in the photosphere. The stars below can be found in the Catalogue of Ap, HgMn and Am stars \citep{renson}. For relatively slowly-rotating stars where the radius that would be measured (from an SED fitting angular diameter and a distance) is a close approximation to the equatorial radius, a reliable rotation frequency can be used to determine the equatorial rotation speed ($v_{\rm rot} = 2\pi R_{\rm eq} f_{\rm rot}$).  With an estimated mass (assisted by the $\tau^1$ Aqr values), the fraction of critical rotation $\omega$ is determined well. 

{\it 29 Psc:} This star's position at the main sequence turnoff gives it value in constraining the age of Psc-Eri. 29 Psc is a known chemically peculiar star (CP4, with weak He and strong Mn), and as a result, it is likely to be a genuine slow-rotator. It was identified as having only a weak rotational signature in \citet{STEREO} though, so knowledge of the rotation frequency $\omega$ is weaker. 29 Psc is a southern spectrophotometry standard star \citep{hamuy,Rubin} that also has a high signal spectrum in the HST Low-Resolution Stellar Library \citep{Pal}, and these data have been used in our SED fits. While there is a wealth of data for the star, systematic effects of the chemical peculiarity probably dominate the uncertainties for the fitted characteristics. That said, our SED fit is in very good agreement with the photometric temperatures calibrated for chemically peculiar stars in \citet{netopil}.

{\it $\nu$ For:} This star is also found at the main sequence turnoff. It is a magnetic Si star, and an $\alpha^2$ CVn variable star with a photometric rotation period of 1.892 d \citep{leone}. Alongside the low projected rotation speed and a radius estimate, the relatively long rotation period supports a truly slow rotation and also allows the spin inclination to be calculated as $36\pm9$ deg. \citet{BESOspect} identify it as a double-line spectroscopic binary based on 18 spectra, but as far as we are aware, this is the only published evidence for a companion. For example, Gaia astrometric indicators like RUWE register as normal. From our SED fits, we compute a larger-than-expected radius, which may reflect a companion contributing significant light. The possibility of determining a binary orbit should be pursued though.

{\it 108 Aqr:} The star has a measured photometric variation period of 3.73 d \citep{mw71,north} of the $\alpha^2$ CVn type, and once again the spin inclination can be estimated ($61\pm20$ deg). Radial velocity observations have found no variation \citep{aands}.


{\it 26 Psc:} This star has a chemical peculiarity designation marked as doubtful in \citet{renson}, but has a detectable photometric variation with a period of 2.534 d \citep{STEREO} that may reflect rotationally-modulated spotting.
26 Psc was found to have constant radial velocity from 11 spectra \citep{BESOspect}, and there are no known co-moving stars.
It was detected in X-rays by EXOSAT \citep{EXOSAT} though, which might be an indication of a later type companion.

\end{document}